# Physics-informed machine learning for composition – process – property design: shape memory alloy demonstration

Sen Liu[1], Branden B. Kappes[1], Behnam Amin-ahmadi[1], Othmane Benafan[2],

Xiaoli Zhang[1]*, Aaron P. Stebner[1,3]*

**Affiliations:**

[1]Mechanical Engineering, Colorado School of Mines, Golden, CO 80401 USA.

[2]Materials and Structures Division, NASA Glenn Research Center, Cleveland, OH 44135 USA.

[3]Mechanical Engineering & Materials Science and Engineering, Georgia Institute of Technology, Atlanta, GA 30332 USA.

*Corresponding authors: aaron.stebner@gatech.edu; xlzhang@mines.edu

**Abstract:** Machine learning (ML) is shown to predict new alloys and their performances in a high dimensional, multiple-target-property design space that considers chemistry, multi-step processing routes, and characterization methodology variations. A physics-informed featured engineering approach is shown to enable otherwise poorly performing ML models to perform well with the same data. Specifically, previously engineered elemental features based on alloy chemistries are combined with newly engineered heat treatment process features. The new features result from first transforming the heat treatment parameter data as it was previously recorded using nonlinear mathematical relationships known to describe the thermodynamics and kinetics of phase transformations in alloys. The ability of the ML model to be used for predictive design is validated using blind predictions. Composition - process - property relationships for thermal hysteresis of shape memory alloys (SMAs) with complex microstructures created via multiple melting-homogenization-solutionization-precipitation processing stage variations are captured, in addition to the mean transformation temperatures of the SMAs. The quantitative models of hysteresis exhibited by such highly processed alloys demonstrate the ability for ML models to design for physical complexities that have challenged physics-based modeling approaches for decades.

**Main Text:**

# 1. Introduction

## 1.1 State of the Art of Data Driven Alloy Design

Predictive design of the process-structure-property-performance (PSPP) materials paradigm [1] is time-consuming for alloys due to the high-dimensional design space and governing physics that span length scales from $10^{-10}$–$10^{0}$ m, the length scale of atomic bonds to metallic components, and time scales of $10^{-14}$–$10^{7}$ s, the time scale of atomic vibrations to aging and corrosion. Decades of global research and development initiatives such as Integrated Computational Materials Engineering (ICME) [2,3] and the Materials Genome Initiative (MGI) [4] have demonstrated the ability for both physics-based and data-driven computations to accelerate the discovery and deployment of new alloys. It has been established that machine learning (ML) can model PSPP relationships of alloys [5,6]. Of equal or greater impact, ML can greatly reduce the number of physics-based experiments and calculations needed to discover and design optimized, new materials [7-9]. This work builds upon these developments to accelerate multi-objective performance design for alloys.

The formulation of effective data descriptors, or “feature engineering,” has emerged as a critical data pre-processing step to improve ML model performance in the field of Materials Informatics. Most such studies have focused on formulating chemical element descriptors to mine large numbers of data curated from high-throughput physics-based calculations [7,9]. For example, density functional theory (DFT) calculations have been used to populate large, organized, and indexed materials databases such as AFLOW [10], OQMD [11], and materialsproject.org [12], which are then mined to build ML models capable of predicting the properties of new single phase materials. The development of data descriptors has been the key to data-driven models to predict the glass-forming ability of metallic glasses [13], band gap energies of thermoelectrics [14], formation enthalpies of semiconductors [15], properties of inorganic crystals [16], critical temperatures of superconductors [17,18], and the structures and band gaps of both Heusler compounds [19,20] and perovskites [21]. Recently, the formulation and integration of new thermodynamic descriptors that consider both entropy and enthalpy, such as the “entropy density of states,” has led to breakthroughs in the discovery of ultra-high temperature ceramics [22]. In addition to DFT databases, these efforts also incorporate large amounts of inexpensive thermodynamic calculations made using CALculation of PHAse Diagrams (CALPHAD) approaches. While these methodologies demonstrate the promise of data-driven ML for materials design, hundreds-of-thousands of predictably formatted data points were available to mine and the cost to generate new data (i.e., run additional calculations) to fill out the missing design space and verify predictions was low. Furthermore, these predictions are of perfect single phase materials, and so they ignore the impact of process variations and microstructure on the performances of the materials.

A vast frontier of discovery and development remains largely unexplored in moving beyond ML-informed discovery and development of single-phase materials using computational materials databases. Alloys are one material class where this is especially true. Most engineering alloys are composed of three or more elements, with the most prolific engineering alloy class, steels, often having 8 or more critical alloying additions and impurities dictating their behaviors [23]. In the best documented ICME examples of such materials (e.g., [24]), the calculations of composition and thermomechanical post-processing effects of the processing are still manual, hierarchical, and bespoke. Today, DFT calculations of ternary alloys and compounds are limited to tractable calculations in terms of model size and computation time; a calculation of a steel considering all its constituent elements is still a decade or more away from being routine. Furthermore, many alloys behave poorly without thermomechanical post-processing to create complex, multi-phase microstructures. The integration of ML with data to automatically search across a broad PSPP alloy design space is still to be attained.

Use of experimental databases for ML is equally challenging. Repositories of materials data often struggle to accommodate the experiment-to-experiment variations in data structures, ontologies, and methodologies, and as a result, the number of data points for any one composition is usually very limited – on the order of ones to tens. It often takes decades to curate even hundreds of data points on process-property variations of a single alloy family, an investment afforded to only a handful of the most promising candidates of a given alloy class, but even then the amount of data from physical experiments is often insufficient to train the ML algorithms necessary to model complex, nonlinear, multidimensional PSPP relationships.

One of the most recent advancements in using ML for alloy design came from an innovation in physics-informed feature engineering. Specifically, Martin et al. developed quantitative search metrics based on known physics of desired crystallographic relationships between two phases (one ceramic, one alloy) to identify phase pairs that would eliminate cracking exhibited by coarse-grained materials during metals additive manufacturing [25]. This physics-informed, ML-driven data mining methodology was applied to records in crystallographic databases that have been curated for decades and contain thousands of records, and millions of potential phase pairs. Hence, the ML greatly accelerated the search process for candidate phases and ensured that all phase pairs of known crystalline phases were considered. Once candidate pairs were identified, however, the connection to processing was developed through traditional (that is, manual) materials engineering. Hence, the ML was used to automate searches within the "structure" space of the process-structure-property paradigm, but the process-property breakthroughs were then attained without further ML guidance.

Inspired by these previous breakthroughs, we have developed a physics-informed feature engineering approach to enable ML to model multi-objective composition-process-property (CPP) relationships using a limited number of experimental data points (several hundred). Specifically, we show that by using feature engineering practices established by the DFT community to model alloy chemistries, together with a new approach that uses physical models of phase transformations to transform thermomechanical post-processing data, ML can work better for alloy design. We demonstrate the ML framework by verifying its ability to predict new shape memory alloy CPP combinations and verify these experimentally.

### 1.2 State of the Art of Shape Memory Alloy Design

Shape memory alloys (SMAs) provide a challenging test case for developing ML based upon physical experiments, largely because physics-based computational methods are still incomplete in their ability to predict shape memory performances from stoichiometry and processing [26,27]. Physical experiment trial-and-error approaches still largely drive the development of SMAs [28-30]. Furthermore, many SMAs do not exhibit shape memory behaviors at all unless they are thermomechanically post-processed with very specific treatments. The shape memory performances of the best performing SMAs cannot be predicted from chemistry alone. In fact, NiTi, the most prolific SMA to date, exhibits poor shape memory properties without proper thermomechanical post-processing [31,32].

In this work, we focus specifically on thermoelastic SMAs; alloys that recover their shape in response to thermal or mechanical load changes via a reversible martensitic (first-order, diffusionless) phase transformation between high temperature, high symmetry austenite phases and low temperature, low-symmetry martensite phases [31]. Established chemistry-based approaches to tune transformation temperatures (TTs) of SMAs for high and low temperature applications is to alter stoichiometries or introduce new alloying elements; for example, within the range of 50 to 52 at.% Ni in NiTi, 0.1 at.% change in Ni changes transformation temperatures by 20 K [33], while Co, Cr, V, Fe, and Mn can be added to NiTi to lower the TTs [34]. Hf, Pd, Pt, Zr and Au increase transformation temperature [35]. In addition to chemistry, post-processing such as mechanical work and heat treatments may also be used to engineer TTs [31,32].

Hysteresis defines the differences between forward (austenite-to-martensite) and reverse (martensite-to-austenite) transformation temperatures, which drives the thermodynamic efficiency of SMA performances; high hysteresis leads to more efficient dampers, while low hysteresis leads to more efficient actuators. Hysteresis can be tuned with chemistry and thermomechanical processing. In the absence of defects or secondary phases, altering the chemistry to tune the lattice parameters of the austenite and martensite phases such that they can share an undistorted phase boundary reduces the hysteresis [30,36,37]. However, SMAs with low hysteresis and high fatigue lives have also been developed using secondary phases and defects, demonstrating the limitations of our current understanding and models for hysteresis engineering for SMAs [29,38-40]. It is the lack of holistic physical models for engineering hysteresis of SMAs with multiphase microstructures and/or multi-step thermomechanical processing routes that further motivates the desire to use ML as a modeling tool. Humans have strived to develop a comprehensive understanding sufficient to formulate accurate physical models for more than 70 years now.

The first ML efforts for SMA design have largely ignored secondary phases and processing. They have mostly focused only on composition to describe the TT value relationships of perfect, single-phase materials [8,41]. Although single-phase SMAs have received much attention in academia, they have not found commercial success as they are usually limited in their ability to sustain multiple cycles without degradation of their functional performance. One recent work broke through this barrier and demonstrated an ability to use a combination of micromechanical structure-property simulations with ML to capture the effects of $Ni_4Ti_3$ precipitates within commercially successful NiTi bulk materials, including modest hysteresis variations of approximately ± 8 °C, within a highly constrained design space of 50.2 < Ni at.% < 51.2 and 0 to 10 vol.% precipitates [42]. While binary NiTi alloys within this chemistry range are used by industry, they are rarely used without first being heavily cold-worked, as precipitation alone does not provide adequate strength for cyclic applications, even with a 20–30% precipitate phase fraction [31,32]. Furthermore, while this work demonstrated a clear Materials Informatics advancement relative to the previous state of the art, structure-property models cannot directly inform processing. An infinite number of heat treatments can result in 0–10 vol.% precipitates, and not all, nor even most, will perform exactly as simulations predict. The design spaces of commercially viable NiTi-based alloys allow hysteresis variations of 100 °C or more, hysteresis variations across multiple alloy/precipitate types, or precipitate volume fractions in excess of 10% (which are most common in commercial applications and demonstrations). Thus, further innovation is needed to create CPP models sufficient to directly inform a manufacturer in the ways and means to make an alloy that is expected to hit a set of new/desired performance metrics; a model that uses ML, physics, or both.

In recent years, NiTiHf alloys have emerged from the decades-long development of more than 1290 known NiTi-based ternary, quaternary, and quinary SMA compositions [43] as one of the most promising classes of next-generation SMAs. What makes them exceptional is their ability to be strengthened sufficiently to exhibit repeatable functional performances using only thermal post-processing treatments, without mechanical cold work. For this reason, they also make a desirable system advancing ML methods for PSPP alloy design based on physical experiments – mechanical work is not needed to attain application-worthy functional performances, hence reducing the dimension of the process design space that needs to be considered to attain a functional material. However, while single-step heat treatments may be sufficient to evoke shape memory behaviors from alloys containing high (15 – 30 at.%) concentrations of Hf [35,44], we have found that multi-step aging treatments are more effective for compositions containing moderate (3 – 15 at.%) amounts of Hf [45-47]. Therefore, there is a need to consider a multi-stage heat treatment design space together with chemistry and synthesis method.

Just over 200 NiTiHf alloy chemistries have been reported upon to-date including those presented in this work, mostly in compositions with high amounts of Hf [43]. Hence, this class of SMAs has been moderately developed to an extent that it is feasible to consider a data-driven approach to their design. And yet, there are still vast expanses of relatively unexplored design space, leaving plenty of opportunity for new discoveries. Specifically, NiTiHf alloys with a mean transformation temperature below 275 K and low hysteresis have not yet been developed. This mean transformation temperature range is consistent with the range desired for most medical implants. The maximum transformation temperature typically cannot exceed 310 K and is usually targeted to be 275 – 295K. Then, for aeronautics, a total transformation temperature range between ~ 215 K to 275 K would enable an aircraft structure to morph autonomously—that is, without an electrical control system—in the transition from takeoff/landing, where temperatures are usually above 275 K, to cruising at altitudes of 8,000 m or more, where temperatures are usually below 215 K [48]. For aerospace, the ability to actuate at cryogenic temperatures would enable autonomous applications such as heat pipes and self-tracking solar arrays to engage when a device is in the path of the sun (e.g., 400 K on the "bright" side of the moon) and disengage in the shadow of a planetary body (e.g., 40K on the "dark" side of the moon) [49,50]. Binary NiTi alloys, the most developed to date, can attain mean TTs within this range, but not the required hysteresis – mean TT combination [42]. Hence, the desire to discover new low-hysteresis NiTiHf alloys with low TTs motivates the validation design targets in this work.

## 2. Experimental Methodology

In our lab, we have been developing NiTiHf alloys for biomedical applications since 2016, many of which have yet to be published. We used twenty-six of these previously unpublished NiTiHf SMA datasets to augment a database of 528 binary NiTi and ternary NiTiHf datasets collected from literature. These datasets are summarized in Table S1. Additionally, previously unpublished validation datasets of two types have been generated: 1. the four vacuum induction melted (VIM) alloy datasets highlighted in orange at the top of Table S2 were generated within the same time period as the aforementioned 26, but these 4 were withheld from the training and testing database; 2. five new alloys were predicted using the trained and tested ML model (see Section 3.6 for design methodology) – their datasets are highlighted in red in rows 5- 9 of Table S2. One of the alloys was then selected to study the sensitivity of the ML predictions to variations of heat treatment schedules; these variations appear in rows 10 – 12 of Table S2.

These new datasets were generated from vacuum induction melting (VIM) or vacuum arc melting (VAM) ingots from high-purity elemental constituents according to previously documented practices (VIM: [45,46], VAM: [51]). The raw ingots were then homogenized in a vacuum furnace at 1050 $^{o}$C for 24 h and 72 h (for VAM and VIM, respectively), then water quenched (WQ). Specimens for differential scanning calorimetry (DSC) measurements were cut from ingot and then solution treated at 1050 $^{o}$C for 0.5 h in an evacuated quartz tube, followed by water quenching (i.e., "Sol"). Pre-aging heat treatments of 300 $^{o}$C for 12 h, followed by air cooling (AC), and aging treatments of different temperatures and times were used. Some samples were not processed with pre-aging and directly subjected to final aging (denotes Sol + Aging). During preaging and aging heat treatments, the specimens were wrapped with tantalum foils to inhibit oxidation. DSC was performed according to ASTM F2004-17 [52] using a TA Instruments Q100 V9.9 with heating and cooling rates of 10 °C/min and temperature range between -150 °C and 150 °C for three cycles. The third cycle was used to measure the transformation temperatures reported in this work. The transformation temperatures were determined with tangent intersection method as shown in Fig. S1.

## 3. Formulation of the Machine Learning Models

### 3.1 Assessment of the database for machine learning suitability

The process-property training and testing database of NiTi and NiTiHf SMAs used in this work consists of 554 data points collected from 528 previously published and 26 previously unpublished (Table S1) datasets. The database used in this work, including citations linked to the original data sources, is publicly available at citrination.com [53]. Each dataset within the database consists of the 48 scalar inputs indicated in Table 1. We derived two calculated outputs from the transformation temperatures of the alloys. Specifically, for each dataset, the martensite finish ($M_f$) (lowest) and austenite finish ($A_f$) (highest) transformation temperatures were used to calculate thermal hysteresis using the definition of the total transformation temperature range $\Delta T$, and the mean transformation temperature $\bar{T}$ of the load-free, thermal martensitic transformation according to (also see Fig. S1):

$$\Delta T = A_f - M_f \text{ and} \tag{1}$$

$$\bar{T} = (A_f + M_f)/2\,. \tag{2}$$

Note that here, we have chosen to study/model the total transformation temperature range as $\Delta T$ as opposed to the DSC endothermic-to-exothermic peak or midpoint-to-midpoint hysteresis definitions, as are often used (e.g., [8,29,30,42]). This choice was made to construct a model that best informs practical uses of SMAs, such as those briefly summarized in Section 1.2, which are limited by the total transformation temperature range, not intermediate differences. While the majority of the transformation temperatures were assessed using DSC data according to ASTM F2004-17 [52], without an applied external stress, in a few cases other methods, such as constant force thermal cycling [54], were used, which can lead to interpretation differences of transformation temperature properties [52,55,56]. These characterization variations are categorized within the database, with the applied stress being assigned a categorical type (1 = tension, 2 = compression, and 3 = zero force) and the stress magnitude (inputs 47–48 of Table 1).

The Ni composition of the alloys within the database ranges from 48.5 to 51.5 at.% while the Hf content ranges from 0 to 30 at.%, as shown in Figs. 1(A, B). The distributions of $\bar{T}$ and $\Delta T$ are shown with respect to Ni vs. Hf content variations in Figs. 1(B, E) and Hf content vs. processing steps in Figs. 1(C, F), respectively. The histograms shown in Figs. 1(A, D) indicate the distributions of $\bar{T}$ and $\Delta T$ for three subcategories of the database that is used in this work: 1) binary NiTi (Hf = 0 at.%), 2) Hf-high (Hf > 10 at.%), 3) Hf-low (0 < Hf ≤ 10 at.%) alloys. The number of Hf-high alloys (369 data sets) is significantly greater than binary NiTi (132 data sets) and Hf-low alloys (53 data sets). Note that all 26 of the previously unpublished datasets (Table S1) are categorized as Hf-low.

Table 1. The input features generated for each dataset through physics-informed feature engineering approaches.

| Feature category | Feature symbol | Feature description | Feature category | Feature symbol | Feature description |
|---|---|---|---|---|---|
| **Elemental properties** | $Z$ | 1. Atomic number | | $q_f$ | 25. Average of energy levels for *f* orbitals |
| | *Gro.* | 2. Periodic table column | | $n_f$ | 26. Average of valence electrons from *f* orbitals |
| | *Row.* | 3. Periodic table row | | $\overline{n_s}$ | 27. Average of *s* unfilled electrons in *s* orbitals |
| | $M_a$ | 4. Relative atomic mass | | $\overline{n_p}$ | 28. Average of *p* unfilled electrons in *p* orbitals |
| | $MN$ | 5. Mendeleev number | | $\overline{n_d}$ | 29. Average of *d* unfilled electrons in *d* orbitals |
| | $r_{cal}$ | 6. Calculated atomic radius | | $\overline{n_f}$ | 30. Average of *f* unfilled electrons in *f* orbitals |
| | $r_{cov}$ | 7. Covalent radius | **Compositions** | [Ni] | 31. Nickel (atomic %) |
| **Reactivities** | $e$ | 8. Valence | | [Ti] | 32. Titanium (atomic %) |
| | $E_{ea}$ | 9. Electron affinity | | [Hf] | 33. Hafnium (atomic %) |
| | $E_i$ | 10. Ionization energy | **Processes variables** | *Syn.* | 34. Synthesis ways (Syn.) |
| | $\chi$ | 11. Electronegativity Pauling | | $T_{sol}$ | 35. Solution temperature |
| **Thermal properties** | $k$ | 12. Thermal conductivity | | $\phi(T_{sol})$ | 36. Transformed solution temperature |
| | $\rho$ | 13. Electrical conductivity | | $t_{sol}$ | 37. Solution time |
| | $\Delta H_{fus}$ | 14. Heat of fusion | | $\phi(t_{sol})$ | 38. Transformed solution time |
| | $\Delta H_{vap}$ | 15. Heat of vaporization | | $T_{pre}$ | 39. Pre-aging temperature |
| | $T_m$ | 16. Melting point | | $\phi(T_{pre})$ | 40. Transformed pre-aging temperature |
| | $T_b$ | 17. Boiling point | | $t_{pre}$ | 41. Pre-aging time |
| **Electronic structure configurations** | $n$ | 18. Total valence electrons | | $\phi(t_{pre})$ | 42. Transformed pre-aging time |
| | $q_s$ | 19. Average of energy levels for *s* orbitals | | $T_{age}$ | 43. Final-aging temperature |
| | $n_s$ | 20. Average of valence electrons from *s* orbitals | | $\phi(T_{age})$ | 44. Transformed final-aging temperature |
| | $q_p$ | 21. Average of energy levels for *p* orbitals | | $t_{age}$ | 45. Final-aging time |
| | $n_p$ | 22. Average of valence electrons from *p* orbitals | | $\phi(t_{age})$ | 46. Transformed final-aging time |
| | $q_d$ | 23. Average of energy levels for *d* orbitals | **Characterization variables** | $\boldsymbol{\sigma_{type}}$ | 47. Applied stress type |
| | $n_d$ | 24. Average of valence electrons from *d* orbitals | | $\lvert\boldsymbol{\sigma}\rvert$ | 48. Applied stress magnitude |

Figs. 1(B, E) show data distributions for Ni, Hf compositions against $\bar{T}$, indicating that $\bar{T}$ generally increases with Hf content and decreases with Ni content, as expected [29,33,35,43,57]. Hf-high alloys have $\bar{T}$ in the range of 224–815 K, Hf-low alloys span 183–395 K, and binary NiTi are in 166–354 K. Fig. 1E indicates that $\Delta T$ is scattered about 25–200 K with respect to Ni, Hf content; hysteresis does not uniquely correlate with composition variations when process

variations also exist within the database, as expected [29]. Fig. 1C shows the variations of $\bar{T}$ as a function of Hf content and heat treatment (HT) variations. The HT variations are categorized as: 1. As-Melted and homogenized (As-Melted), 2. solid solution annealed following homogenization (Solutionized, Sol), 3. directly aged from homogenization (Direct Aged), 4. melted, solid solution annealed, and then aged (Sol + Aged), and 5. solid solution annealed, then pre-aged and finally aged (Sol + Pre-Aged + Aged). The physical reasons for investigating these different heat treatment strategies in processing NiTiHf alloys are well established, e.g. [45,46]. In examining Fig. 1C, it is obvious that at specific Hf content, $\bar{T}$ values are greatly scattered due to variations of Ni content and processing. Still, across different Hf content values, $\bar{T}$ shows a generally increasing for Hf > 10 at.%. Fig. 1F shows the analogous dependence of $\Delta T$ as a function of Hf content and HT variations. The maximum observed $\Delta T$ within this dataset increases with Hf content, though the means and modes of different Hf contents do not show such obvious variation. Furthermore, different synthesis methods were used, which are also known to influence transformation temperature [58,59]. These methods were categorized as vacuum induction melting, vacuum arc melting, and other.

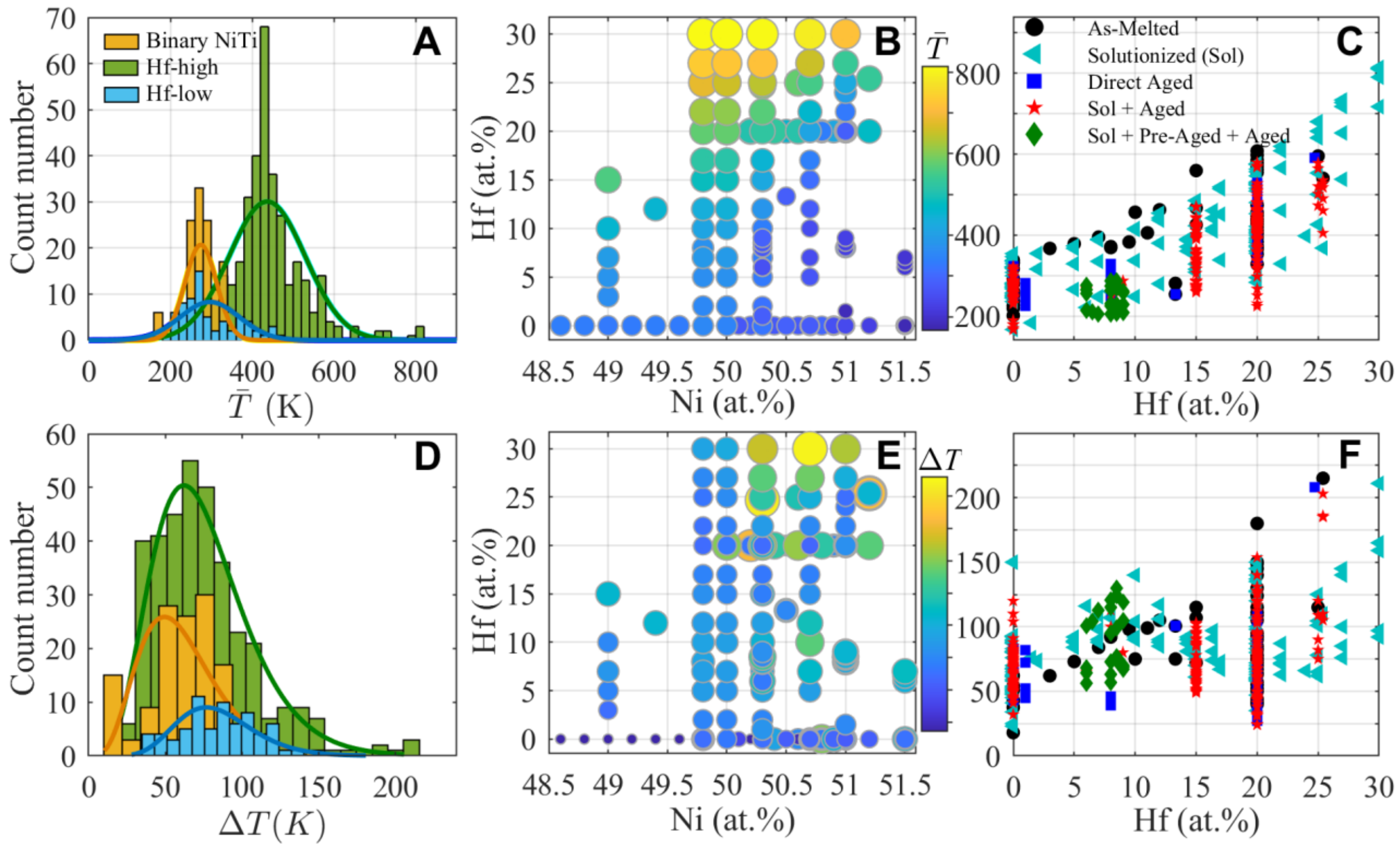

**Figure 1. Visualization of the database for composition – process – property relations.** Histograms indicate the distribution of properties (A) $\bar{T}$ (mean transformation temperature) and (D) $\triangle T$ (transformation temperature range) for three subclasses of NiTi and NiTiHf alloys within the database. The dependence of (B) $\bar{T}$ and (E) $\triangle T$ on Ni and Hf content variations; larger marker sizes indicate higher property values and vice versa. The variation of (C) $\bar{T}$ and (F) $\triangle T$ as a function of Hf content and processing path variations. The five process path categories are further described in the Section 2.1.

Altogether, the analyses of the database given in Fig. 1 shows that both composition and the process variations significantly impact $\bar{T}$ and $\Delta T$, indicating the correlation among composition, process and resultant $\bar{T}$ and $\Delta T$ exists such that the ML model is suitable to capture this correlation. Furthermore, no obvious empirical model that describes the composition-process-property relationships of NiTi and NiTiHf SMAs is evident from 3D visualizations, indicating that the relationship correlations are of higher order. Still, even these lower dimensional visualizations show a clear trend across the input vs. output design space, indicating that statistical approaches are suitable to model these relationships. Machine learning is therefore used as the modeling approach to describe these relationship correlations.

### 3.2 Machine learning algorithm selection and tuning

A regression model is required to estimate the continuously valued transformation temperature metrics ( $\bar{T}$ and $\Delta T$) from composition and processing inputs, and so we fit the data using Support Vector Regression (SVR), random forest (RF), Gaussian Process Regression (GPR) models [60]. SVR and RF model performances were optimized using standard practices, including grid search and cross-validation for hyperparameter optimization. However, both the superior prediction accuracy of GPR and its ability to estimate not only the response, but also the variance in the response made this model more suitable for extrapolation near the supported domain; that is, estimating transformation temperature metrics just beyond the composition and processing space spanned by the training set.

For GPR, the choice of covariance function captures what is known about the relationship between observations; that is, whether the system is periodic, decaying, etc. With no *a priori* knowledge about the relationships that exist between composition, processing and transformation temperature, the isotropic squared exponential covariance function was used.

GPR differs from other ML algorithms in that, except for the choice of covariance function, the model hyperparameters are fit from the input data through a maximization of the log marginal likelihood of the model hyperparameters. Thus, the bias-variance tradeoff is made automatically as part of model training. Although the ability to fit the model hyperparameters directly from the data obviates the need for the typical model hyperparameter grid search used in SVR and RF, the performance of the model is not necessarily concave, or even well-behaved, in the space spanned by the covariance function model parameters. Therefore, GPR models must be optimized by restarting the model training at different locations in the hyperparameter space. To address this issue, the model is restarted in decade steps from -1000 to 1000 for each of the isotropic squared exponential covariance function hyperparameters, $\theta = \{\sigma_n^2, \sigma_f^2, l\}$. Matlab GPML version 4.2 [61] was used to implement the GPR models.

Cross-validation (CV) is used to guard against overfitting by estimating the out-of-sample error. This error was estimated from ten-fold, three-fold, and leave-one-out cross-validation [62] using standard performance metrics: coefficient of determination ($R^2$), root-mean-square error (RMSE), mean-absolute error (MAE), and mean relative error (MRE). Ten-fold cross validation using RMSE as the performance metric resulted in the most accurate model.

### 3.3 Physics-informed feature engineering

A previously established feature augmentation methodology for material chemistries uses elemental property attributes, electronic structure attributes, crystal structure representations, and density functional theory (DFT) calculated formation energies to augments material compositions [14]. We used this established approach to generate additional inputs (1 – 30 in Table 1) based upon the physics of alloy compositions (31 – 33 in Table 1). Each augmented composition feature $A_i$ for $i = 1, \dots, 30$ as indicated in Table 1 was calculated as the weighted fraction $f_x$ of each constituent element $x = \mathrm{Ni, Ti, Hf}$ according to Eq. (3).

$$A_i = A_{\mathrm{Ni}} f_{\mathrm{Ni}} + A_{\mathrm{Ti}} f_{\mathrm{Ti}} + A_{\mathrm{Hf}} f_{\mathrm{Hf}} \tag{3}$$

Values for $A_x$ such as elemental features, electronegativity $\chi$, melting point $T_m$ and valence electrons *n* were taken following the same procedure and the data sources as [14]. Other augmented composition features—such as heat of fusion $\Delta H_{\mathrm{fus}}$, electron affinity $E_{\mathrm{ea}}$, ionization energy $E_{\mathrm{i}}$, thermal conductivity *k* or valence energy levels *q*—were taken from [63-66].

As is shown in Section 4.1, a model using only these element-based features shows high uncertainty for both $\bar{T}$ and $\Delta T$ and poor ability to model $\Delta T$, which was not surprising considering that it is well established that process variations strongly impact the transformation temperature properties of SMAs, as previously documented for this database in Section 3.1 and Fig. 1. Hence, we incorporated the process (34-46) and characterization (47-48) data features in Table 1 into the database.

For heat treatment features, we initially used heat treatment times and temperatures as they were entered in logbooks (the *T* and *t* features: 35, 37, 39, 41, 43, 45 in Table 1). However, as is also shown in Section 4.2, we found that the uncertainties of the ML predictions were still characterized by standard deviations of 100 K or more. Ultimately, we realized that mathematical functions known to model the physics of the kinetics of the solid solution and precipitation phase transformations that result from the heat treatments are highly nonlinear. For precipitation, for example, there are temperatures below which no transformation occurs, and temperatures above which transformation is complete. This points toward a sigmoidal relationship between precipitation temperature and transformation, a relationship described empirically by the logistic sigmoid. Thus, by applying such functions to the heat treatment times and temperatures (the $\phi(T)$ and $\phi(t)$ features 36, 38, 40, 42, 44, 46 in Table 1), ML models could be informed *a priori* of physiochemical knowledge of mathematical nonlinearities, allowing a relatively inexpensive machine learning regression algorithm, like GPR, to make more reliable predictions with limited amounts of training data, as we proceed to demonstrate.

The functions $\phi$ used to transform the heat treatment (HT) times and temperatures were determined from known empirical models. Specifically, the JMAK (Johnson-Mehl-Avrami-Kolmogorov) growth kinetics model [67-70] gives the relation between the fraction of phase transformed material *Y*, relative to the time, *t* according to:

$$Y(t;T) = 1 - e^{Kt^n} \tag{4}$$

where *K* is a temperature-dependent growth constant and *n* describes the orders of the growth. The above function can be converted to $\ln\big(ln(1-Y)\big) = \ln K + n \ln t$. Since *K* and *n* are constant, the phase transformed fraction can be expressed linearly as a function of ln(*t*). Therefore, $\phi$(t) = ln(t) was used to transform the heat treatment time features.

Similarly, a phase transformation temperature $\theta$ may be related to precipitate growth using a sigmoid function as in Eq. (5). This formulation reflects that at insufficiently low temperatures, there is an effectively zero probability ($\sigma$) of precipitation, while at higher temperatures, precipitation is very likely to have occurred. In this work, $\theta_{sol}$ = 850 °C was used for all alloys. When the Hf content was less than 3 at.%, $\theta_{pre}$ and $\theta_{age}$ were set to 200 and 400 °C to model $Ni_4Ti_3$ precipitation kinetics [31]; for Hf content large than 3 at.%, $\theta_{pre}$ and $\theta_{age}$ were set to 300 and 500 °C, respectively, to model H-phase precipitation kinetics [44-46].

$$\sigma(T) = \frac{T}{1 + e^{\theta_i - T}} \tag{5}$$

Again, Table 1 summarizes both the recorded and engineered input features.

### 3.4 Feature down-selection

Feature engineering expanded the number of model inputs to 48 (Table 1); the total number of observations is 554. By increasing the number of features, feature engineering increases the dimensionality of the problem, leading to the “curse of dimensionality” [71]. The objective function in most—virtually all—deep learning algorithms is pathological: it is non-convex with many local minima in the space spanned by the model parameters [72]. GPR models suffer from an analogous pathology. Escaping the local minima in the model parameter space requires that the model be restarted repeatedly at reasonable initial parameterizations of the GPR model priors to find the optimum parameter posteriors [73]. Following from the manifold hypothesis, as the dimensionality of the problem increases, so does the number of local minima. The manifold hypothesis asserts that high dimensional data generally lies in the vicinity of a lower dimensional manifold [74,75]. More concretely, if the solution of a problem lies on the surface of a 3D cylinder, $y = f(x, y, z)$, then the solution may be better represented in 2D, $y = f(\theta, z)$, that is, on the “unrolled” cylinder. Solving the problem on the lower dimensional surface requires fewer observations—there are fewer $\theta, z$ combinations than $x, y, z$ combinations in the problem domain—and is less prone to overfitting since many of the local minima in the model parameter space lie in the error hyperplane (in the example above, the direction normal to the cylinder surface—the radial direction). Therefore, it is generally beneficial to reduce the dimensionality of the problem to eliminate nuisance variables, improve model performance, and reduce the number of necessary training observations.

Here, the relative importance of the features in determining each $\bar{T}$ and $\Delta T$ were ranked using the mutual information (MI) score method [76], while redundancy was evaluated using Pearson correlations [77]. The composition feature subset was evaluated independent of the process and characterization features, primarily for the purpose of evaluating the contribution of the new feature additions in this work relative to the previous state of the art.

The Scikit-learn python implementation of these algorithms were used [78]. The MI score $I(X, Y)$ from input feature $X$ and output property $Y$, can be computed from,

$$I(X,Y) = \iint p_{xy}(x,y) \log \frac{p_{XY}(x,y)}{p_X(x) p_Y(y)} dx\, dy \quad (6)$$

where $p_{XY}$ is the joint probability density function of X and Y, and $p_X$ and $p_Y$ are the marginal probability density functions. This integral equal zero if and only if the feature ($X$) and property ($Y$) are independent, and a higher MI score indicates greater dependency. Correspondingly, Pearson correlation between feature pairs $r_{x_{ij}}$ or feature and property $r_{xy}$ uses the standard definition,

$$r_{xy} = \frac{\sum_{i=1}^{n}(x_i - \bar{x})(y_i - \bar{y})}{\sqrt{\sum_{i=1}^{n}(x_i - \bar{x})^2} \sqrt{\sum_{i=1}^{n}(y_i - \bar{y})^2}} \quad (7)$$

where $n$ is sample size, $x_i$ and $y_i$ are the individual sample points and $\bar{x}$ and $\bar{y}$ are the sample means.

The results for the composition features are shown in Figs. 2(A, B). The electronic structure attributes $n_s$, $q_p$, $n_p$, $\overline{n_s}$, $\overline{n_p}$, $\overline{n_f}$ exhibit very low MI scores and were removed. Then, augmented composition feature pairs with correlation coefficient larger than 0.90 (Fig. 2B) were taken to be highly correlated, hence redundant and removed: $\rho$, $\triangle H_{fus}$, $\triangle H_{vap}$, $T_m$, $T_b$. The most important (assessed via MI scores) variables of the subsets of highly correlated augmented composition features were retained. The raw composition features (31-33 in Table 1) were considered separately. Ti and Hf at.% were removed since they are inversely correlated with many of the augmented composition features. Though its importance is marginal based strictly on MI score, Ni at.% was retained since it is only weakly correlated with most other features. The Pearson correlation matrix of the final down-selected set of 11 composition features is shown in Fig. 2C.

The MI scores of process features are generally lower than the most important composition features, yet they are not negligible (Fig. 2D). The applied stress and pre-aging features have a relative low MI score, likely due to insufficient statistical sampling: transformation temperatures are determined using DSC (stress-free) in most cases and pre-aging is a rare post-processing strategy. This sample size sensitivity was not a concern in the composition-based feature down-selection exercise since these features were equally represented (i.e., fully dense) across the database. The Pearson correlations between input process features and each output property in Fig. 2E show that preaging and applied stress correlations are of comparable importance as the other process variables, even though they have low MI scores. Hence, these features were retained since the reason for low MI scores is not a lack of importance, but rather, the sparsity of samples in these combinations of conditions. Fig. S2 further demonstrates the that a high MI score does not necessarily indicate a strong correlation and vice-versa for 10 of the down-selected composition features (correlations for the 11th, Ni at.%, are documented in Fig. 1). Unlike the composition-based features, the general types of process features are not expected to be cross-correlated; solid solution annealing dissolves precipitates while aging forms them – the physics are different. However, the untransformed vs. transformed HT times and temperatures for each heat treatment path should be highly correlated, since the inputs are the same, they are just operated on by different functions (unity multiplication in the untransformed case). Hence, cross-correlation analysis to down-select the process and characterization features was not performed; instead, model performances were examined using the transformed vs. untransformed HT feature subsets.

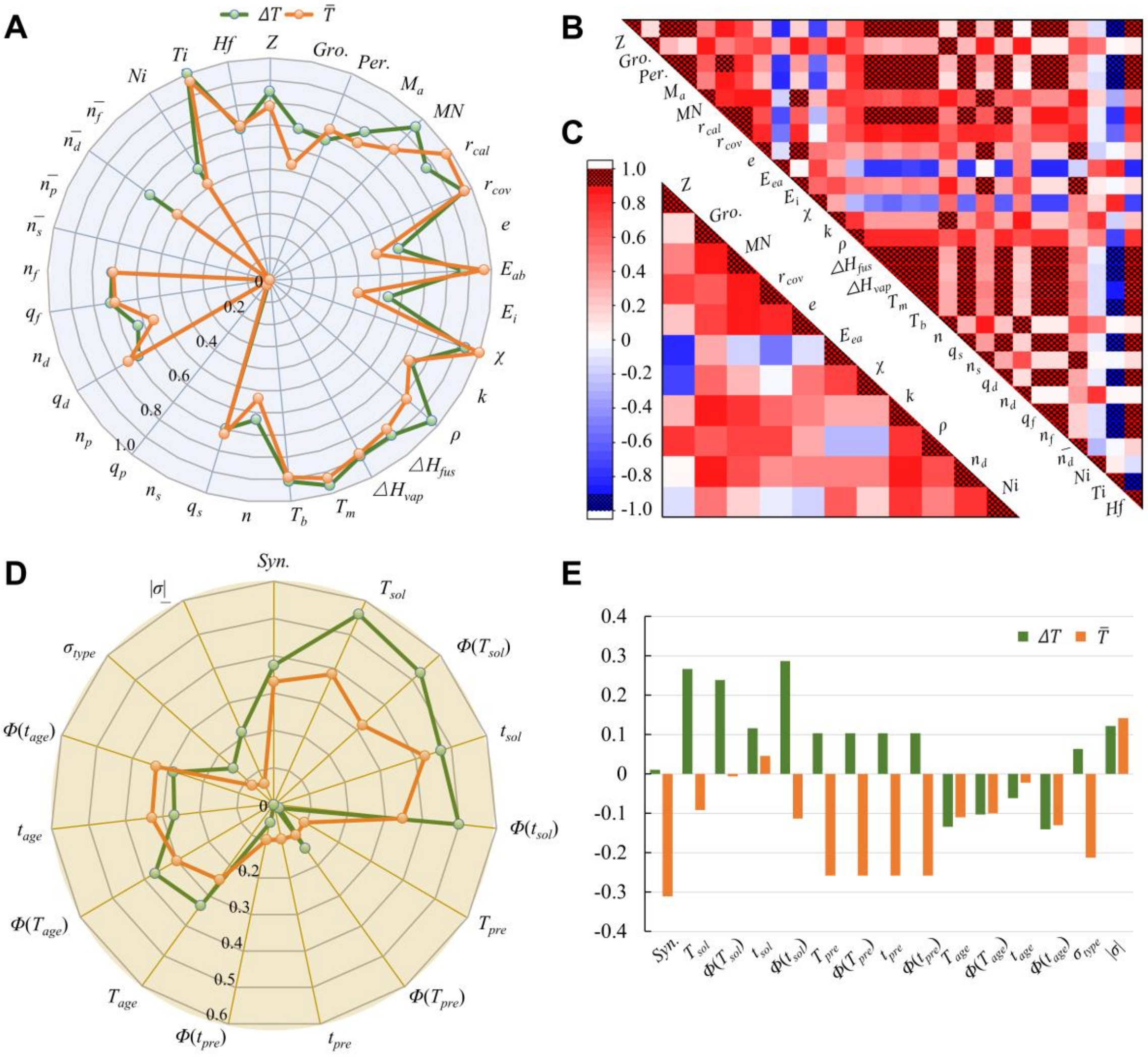

**Figure 2. Feature down-selection criterion.** Mutual information (MI) score spider plots indicate the overall significance of composition (A) and process (D) features in determining each output property $\bar{T}$ and $\Delta T$. Pearson cross-correlation matrices indicate the relative redundancies of (B) all composition features and (C) the down-selected set of composition features used for ML modeling.

Pearson correlations between process and characterization input features and output properties (E) confirm that even though pre-aging and applied stress showed low MI scores in (D), their correlations to the outputs of interest are of the same magnitude as the other process variables.

### 3.5 Alloy design prediction methodology

Twenty-seven thousand transformation temperature and hysteresis estimates were made using the trained ML models, exhaustively exploring the 4,500-point compositional design space ($0 \leq$ Hf $\leq 30$ at.% and $49 \leq$ Ni $\leq 52$ at.% at step sizes of 0.1 at.% Ni and 0.2 at.% Hf), 3-point annealing space (*Sol* (1050 °C/0.5h, WQ), *Sol* + *Aged* (1050 °C/0.5h, WQ + 550 °C/3.5h, AQ), and *Sol* + *Pre-Aged* + *Aged* (1050 °C/0.5h, WQ + 300 °C/12h, AQ + 550 °C/3.5h, AQ) heat treatment paths), and 2-point synthesis space (vacuum induction melting, VIM, or vacuum arc melting, VAM). Ternary diagrams visualizing the resulting calculations are given in Fig. S5 for the mean $\bar{T}$ and $\Delta T$ predictions, $\mu$, and Fig. S6 for the standard deviations, $\sigma$. These predictions were then sorted according to minimum $\Delta T$ and filtered to identify estimates with a $\bar{T}$ between 230 K and 260 K. From this filtered set, seven composition-process combinations that were unique from each other by at least 1 at.% Hf or 0.1 at.% Ni were identified from the VAM predictions for experimental synthesis and characterization (the available VAM furnace was able to make smaller ingots more quickly, saving cost and time, while still sufficiently testing the utility of the ML model).

These predictions can also be mined for physical insights – while not the focus of the main article, this model capability is demonstrated in the captions of Fig. S5, studying VAM vs. VIM processing effects, and Figs. S4 & S7, studying composition and heat treatment path effects.

## 4. Machine Learning Model Assessments

To quantify the impact of the new physics-informed features, we proceed to evaluate them against the previous state of the art (e.g., [8,41]) in Fig. 3. We also evaluated a physics-inspired approach of training and testing using different test-train combinations in the spirit of leave-one-out cross validation [62]. In this approach the binary NiTi, Hf-high, and Hf-low data subsets of Fig. 1A act as the instances; that is, a model trained on NiTi and Hf-high is tested on Hf-low, a model trained on NiTi and Hf-low is tested on Hf-high, and a model trained on Hf-low and Hf-high is tested on NiTi. The results are given in Fig. S3 and, as discussed in that caption, demonstrate that while such an approach may not perform well in a statistical sense, it can be a useful tool for understanding comparability across instances that express different physics, or for establishing expectations of how well an ML model trained on one set of alloys could work to predict new alloys.

### 4.1 Down-selected composition feature models

The model performance fitted on elemental composition features are evaluated in Fig. 3A ($\bar{T}$) and Fig. 3E ($\Delta T$). An ideal model would place all predicted values on the brown diagonal line. For $\bar{T}$ in Fig. 3A, the model performs decently, consistent with previous work that evaluated the ability for ML to predict a transformation temperature [41], considering that here we are predicting the average of all transformation temperatures, not just a single temperature. The $R^2$=0.83 indicates that the model generally trends with the data, while the distribution of errors histogram in the inset approximately shows that the model predicted 90% of data with less than ±20% relative error. These metrics indicate that the ML model to predict $\bar{T}$ is generally working, but also that there could be room for improvement, especially further considering that the orange and yellow bands represent a mean predicted standard uncertainty $\bar{\sigma}$ of 45 K. Knowing that the transformation temperature of an SMA that is returned by the ML model has a 69% expected accuracy within $\pm$45 K ($\sigma$) and 99% expected accuracy within $\pm$ 90 K ($2\sigma$) is not practically useful - typically for SMA application design, engineers need to know these temperatures to within 5 to 10 degrees.

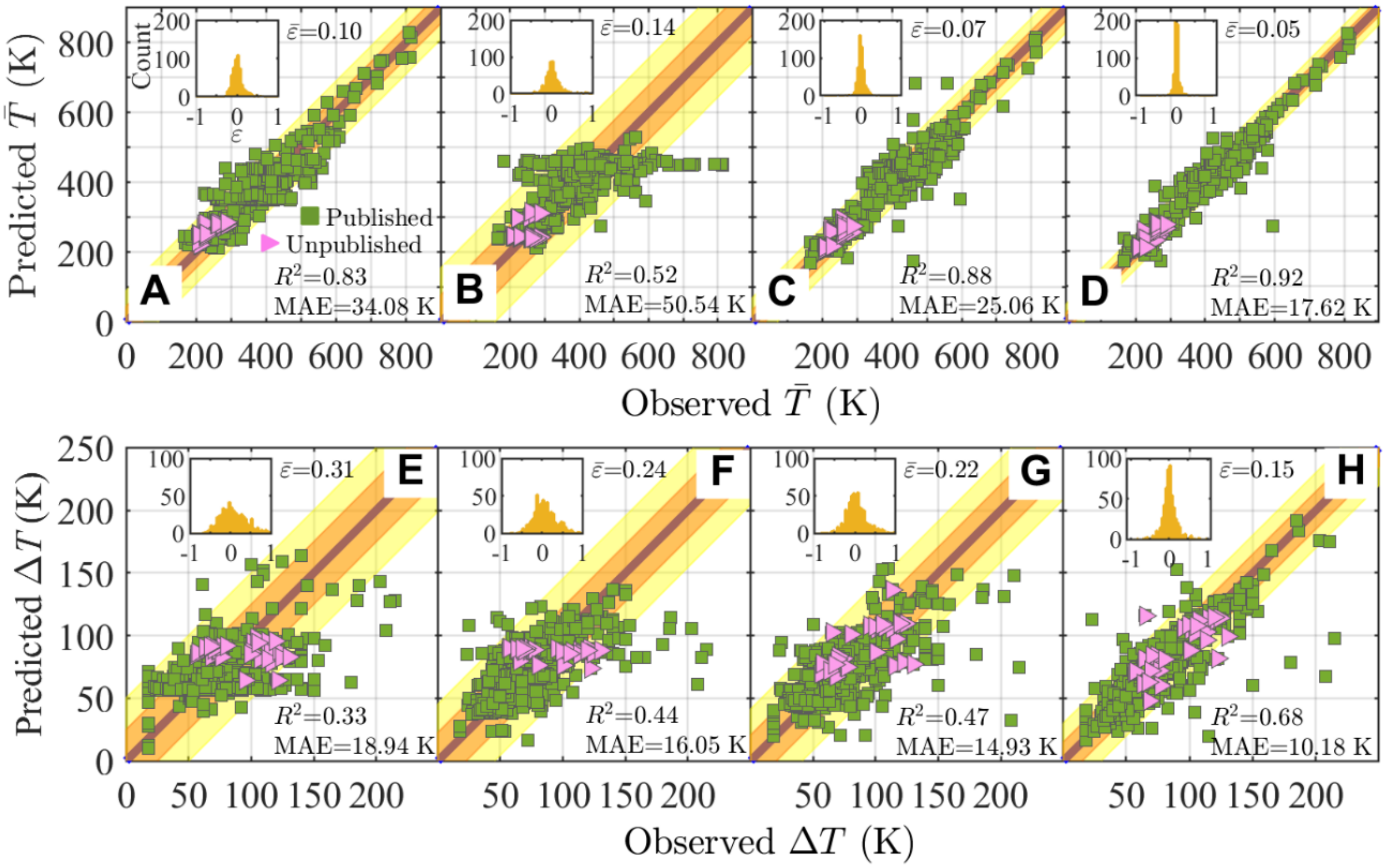

**Figure 3. Machine learning model assessments.** Model predictions with 10-fold cross-validation for (A-D) $\bar{T}$and (E-H) $\triangle T$. (A, E) Trained only with down-selected composition features. (B, F) Trained with untransformed process features. (C, G) Trained with down-selected composition and untransformed process features. (D, H) Trained with down-selected composition and physics-informed transformed process features. The inset histogram indicates relative predicted error $\varepsilon$. Uncertainty is represented by bands colored at $\pm\sigma$ (orange) and $\pm 2\sigma$ (yellow) about the theoretically perfect predicted vs. observed trend line (brown), where $\sigma$ is the standard deviation across all predicted vs. observed differences for each model. $R^2$, MAE, and mean error $\bar{\varepsilon}$ values are also given for each model. Published data are indicated with green squares and previously unpublished data with pink triangles.

The need for a better model becomes more evident in considering the $\Delta T$ cross-validation (Fig. 3E). In regards to predicting thermal hysteresis (here defined to be the total transformation temperature range $\Delta T$), an $R^2 = 0.33$ indicates that the model does not work at all, only 46% of the test data were predicted with relative error lower than ±20% and $\bar{\sigma}$ was 25 K. Visually, it is also evident that the predicted vs. measured values do not trend with the brown diagonal line. Physically, hysteresis is as important, if not more important than the mean transformation temperature, as it determines the efficiency of a shape memory device and has also been suggested to be an indicator of damage and fatigue [47,79]. Physically, the failure of the ML model trained only on composition-based data to predict $\Delta T$ is expected. While for a single, homogenous solid solution SMA with fixed processing, hysteresis can correlate with composition changes in the vicinity of the global minimum [28,37], it is well established that in looking more broadly, especially across different alloys and process variations [29] as we have asked the model to do here, hysteresis is not well correlated with composition variations by themselves. This result emphasizes that ML cannot circumvent fundamental physics and statistics; fitting an ML model, just like an empirical model, requires a full complement of those independent variables on which the response depends. Next, we evaluate the impact of adding process features, first in isolation, and then in combination with the composition features.

### 4.2 Process feature models

Process variations strongly impact SMA transformation temperature properties [31], as they do the properties of all alloys [80]. However, cross-validation of the ML models trained only on the untransformed process features indicates poorer performance in predicting $\bar{T}$ (Fig. 3B) and slightly improved, but still poor, performance in predicting $\Delta T$ (Fig. 3F) relative to the models trained only on composition features. Here, neither model works well at all, as is visually apparent from the lack of correlation in the predicted vs. measured data with the brown diagonal lines. Again, this is not surprising as more than 80 years of understanding the physical metallurgy of SMAs has established that both composition and processing dictate transformation temperature properties. Again, the ML cannot circumvent fundamental physics and statistics.

### 4.3 Untransformed process features combined models

ML models to predict both $\bar{T}$ (Fig. 3C) and $\Delta T$ (Fig. 3G) are improved by considering composition and processing variations simultaneously, even though we have not yet informed the ML with our physical knowledge of the mathematical non-linearities in the process variations. The $R^2$ value of the $\bar{T}$ prediction model (Fig. 3C) increased to 0.88, while the mean error and MAE decrease to 0.07 and 25 K, respectively. Most notably, the mean uncertainty $\bar{\sigma}$ decreases to 35 K—a 10 K reduction in the variance of the predictions, showing there is a very practical benefit to including both composition and process features. However, there are still noticeable outliers from the desired trend in the cross-validation data, indicating that the ML model has not yet captured all of the high dimension composition-process-property correlations with regard to determining $\bar{T}$, or we have not provided the ML model with all the features that are needed to determine these correlations.

Though improved, the ML predictions of $\Delta T$ are still poor (Fig. 3G). Visually, the data are trending better with the brown diagonal line, but the $R^2$ value of 0.47, while improved from 0.33 (Fig. 3E), is still far from ideal ($R^2$=1). This result indicates that it is more challenging to model $\Delta T$ than $\bar{T}$ using ML, which is also consistent with outstanding challenges in understanding the physical mechanisms that determine the hysteresis of SMAs that contain precipitates [29,39].

**4.4 Physics-informed transformed process features combined models**

Informing the ML models of the physical nonlinearities of the heat treatment features further improves predictions of $\bar{T}$ (Fig. 3D) and $\Delta T$ (Fig. 3H). While ML for $\bar{T}$ already performs well (Fig. 3C), the improvement to nearly 92% of the test predictions being made with less than 10% error, visually apparent in the much sharper histogram of errors in the inset, together with another 10 K reduction of $\bar{\sigma}$ to 25 K is impactful. It is also visually noticeable in examining Fig. 3D vs. Fig. 3C that the number of outliers from the desired trend has been obviously reduced – there is now only 1 obvious outlier and a limited number of data points fall outside of the $\pm 2\sigma$ (yellow) interval. Furthermore, the ML model to predict $\Delta T$ (Fig. 3H) has improved to the extent that we can now say it has begun to work, as indicated by an $R^2$ = 0.68, $\bar{\varepsilon}$ = 0.15, MAE = 10 K, and a $\bar{\sigma}$ = 15 K. It is also qualitatively apparent through an obvious sharpening of the histogram of errors together with a much stronger trend with fewer outliers relative to the ideal model (brown diagonal line).

**5. Practical Validation of the Physics-Informed Machine Learning Model**

SMA applications depend on both transformation temperature and hysteresis. And while it is clear from the preceding section that the performance of an ML model improves with more complete information, it is also clear from Fig. 3 that even this model does not capture the complete set of physics governing NiTiHf's thermal transformation. Contrarily, it is not clear how the remaining extrinsic error in the ML model will impact its practical use as a tool to inform the engineering and manufacturing of new alloys to meet application-driven performance metrics. To clarify the utility of the ML model, it is validated using two methods: 1) we reserved four of the previously unpublished NiTiHf datasets for validation (see Section 2 and Table S2), which were not used at any point during training, training, or cross-validation, and 2) we made five new blind predictions (see Table S2) using the trained ML models following the design methodology given in Section 3.5, and then experimentally validated those predictions using the methodology given in Section 2 with no further modifications to or retraining of the ML models.

Fig. 4 shows the validation results within the context of the trained-and-tested models. Specifically, the four unpublished alloys reserved for validation (orange circles) and the five "blind prediction" alloy designs (red triangles) are plotted on top of the training and testing datasets (small white circles), and the $\pm\sigma$ (orange diagonal band) and $\pm 2\sigma$ (yellow diagonal band) uncertainty intervals of the trained models that were determined from cross-validation. Again, recall from Section 3.5 that the design targets for the blind predictions were to achieve a transformation temperature between 230 K and 260 K while minimizing $\Delta T$. This target range is indicated with dashed lines in Fig. 4.

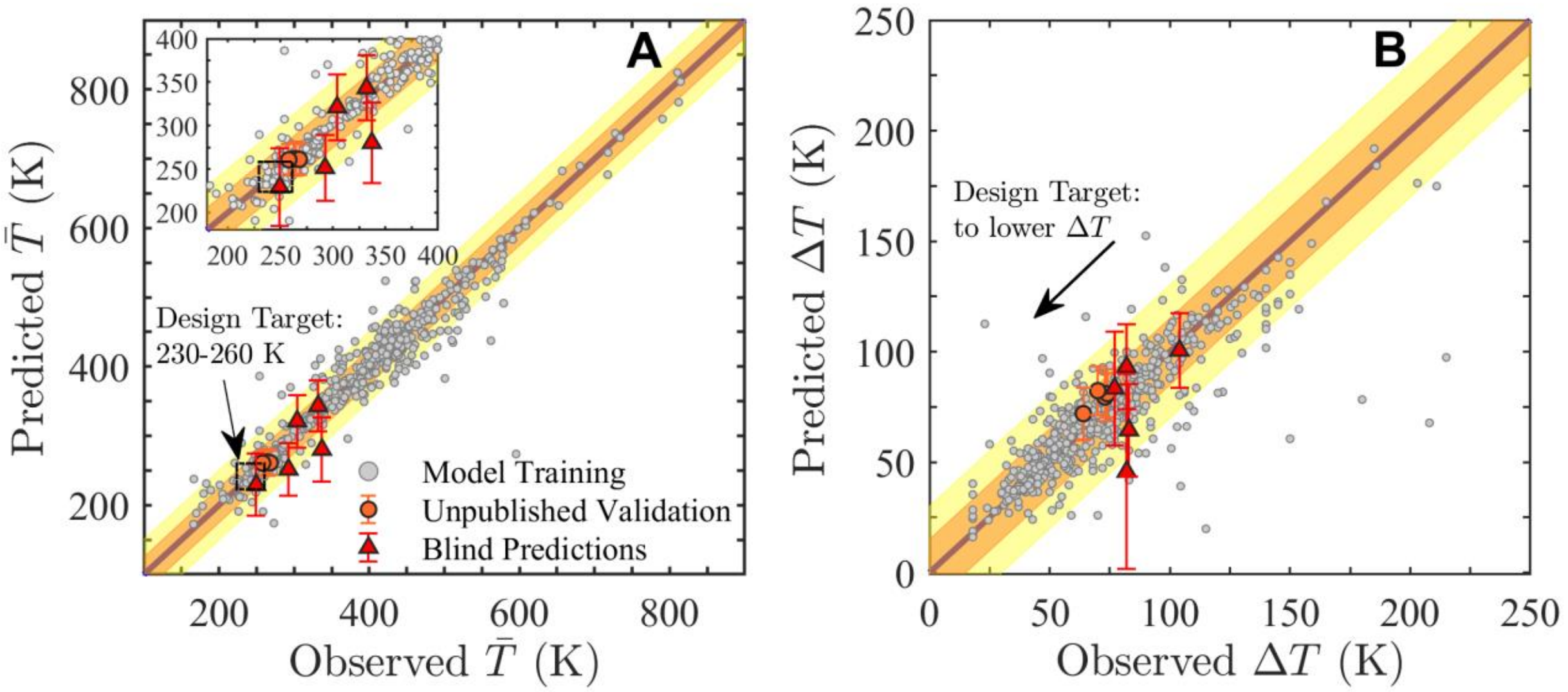

**Figure 4. Experimental validation of the ML models**. The predicted (A) $\bar{T}$ and (B) $\triangle T$ values are plotted against the experimental observations for the validation datasets, including both unpublished data that were reserved from the model training and testing process, as well as the new blind predictions made with the trained and tested models. The error bars indicate the one-sigma prediction uncertainty. These validation data points are plotted over the training data points, as well as the $\pm\sigma$ (orange) and $\pm 2\sigma$ (yellow) uncertainty intervals that resulted from cross-validation of each model. The validation data are summarized in Table S2.

These results validate the performance expectations of the models and their utility for predictive design. Specifically, most experimental verifications of the model estimates lie within $\pm\sigma$ and all but one lie within $\pm 2\sigma$. Only one of the blind alloy composition-process-property predictions was expected to fall within the $\bar{T}$ target window; the $Ni_{50.7}Ti_{46.3}Hf_3$ alloy exhibited $\bar{T}$ = 249.5 K and $\Delta T$ = 77 K, consistent with the 229 ± 45K and 83 ± 26 K predictions. Contrarily, one blind prediction sits on the cusp of each $\pm 2\sigma$ interval for both $\bar{T}$ and $\Delta T$: $Ni_{50}Ti_{47}Hf_3$. While the measured properties of this alloy are $\bar{T}$ = 337 K and $\Delta T$ = 82 K, they are predicted to have a lower transformation temperatures and hysteresis. Physically, this prediction represents an edge case. To form the strengthening H-phase precipitates in NiTiHf alloys using heat treatments, an alloy must have more than 50 at.% Ni in addition to moderate amounts of Hf, so that an excess of both Ni and Hf are available to supply the precipitates [81]. However, this composition is not Ni-rich and, therefore, H-phase precipitation is not expected. Still, nearly all NiTiHf alloy development has occurred since the discovery of H-phase precipitation in the late 1990s [82]; thus, nearly all development has been on Ni-rich compositions, and nearly all of the training data used for ML model development was on Ni-rich compositions. Thus, this poor performance of the ML models for the $Ni_{50}Ti_{47}Hf_3$ alloy is due to the models not having enough data to learn that H-phase precipitation ceases when the Ni-content falls at-or-below 50 at.%. We intentionally made our predictions to Ni-contents as low as 49 at.% to test our edge cases, but in practice, these models should not be expected to perform well for equi-atomic and Ni-lean compositions, considering both physical and training data limitations. The model did alert the user to potential issue with this prediction, as is evident in the abnormally large $\pm\sigma$ error bar associated with this data point in Fig. 4B.

In further examining Fig. 4, the reserved, unpublished data generally perform better than the blind predictions—they exhibit less scatter about the ideal (brown line) response and smaller $\pm\sigma$ error bars. This improved performance is likely because these alloys are most similar to the 26 unpublished data that were used in the training and testing of the models (Table S1). The Ni and Hf at.% contents are shared with other alloys, whereas the "blind predictions" have either more or less Hf than the training data and different amounts of Ni. Thus, the validation analysis has assessed the ability for the model to be used for statistically supported predictions—in this case, the reserved datasets—as well as unsupported predictions—in this case, the blind predictions. In both cases the model is validated, though the poor performance of the $Ni_{50}Ti_{47}Hf_3$ demonstrates that the ability to make unsupported predictions is limited by a physical boundary at Ni-content of 50 at.%. Similarly, there is a likely upper bound on Ni and Hf content at the point where alloys no longer exhibit shape memory behavior, but these edge cases were not reached due to artificial limits placed on the design search.

## 6. Conclusions

Fig. 5 summarizes the impacts of this work. Specifically, the previous state of the art for ML of precipitate strengthened NiTi SMAs resulted in the NiTi Pareto Front [42] that lies between 280 K < $\bar{T}$ < 350 K and 50 K < $\Delta T$ < 60 K. The previous state of the art for developing NiTiHf SMAs is indicated by the Prior NiTiHf Pareto Front and the green square markers. Contextually, it should be noted that until very recently (see Section 2), the majority of NiTiHf SMA development has been driven by ~500 K actuator performance metrics, so this Prior NiTiHf Pareto Front is biased toward $\bar{T}$ higher than the New NiTiHf Pareto Front, which considers previously unpublished NiTiHf alloys driven by medical device performance metrics and blind predictions driven by developing alloys with 230 K < $\bar{T}$ < 260 K and as low $\Delta T$ as physically possible.

Overall, in examining Fig. 5, it is indisputable that the development of NiTiHf SMAs has enabled a much broader, commercially viable SMA design space than was achieved in more than a half century of binary NiTi SMA developments. The NiTiHf Pareto fronts are far more expansive than the NiTi Pareto front. The greatest contribution of the new data presented in this work has been to push the NiTiHf Pareto Front further into the realm of biomedical and aerospace applications, although the pursuit of autonomous aircraft actuation SMAs remains an open challenge. In considering the physically reasoned (i.e., unpublished) NiTiHf alloys vs. the ML-designed (i.e., blind prediction) alloys, the New NiTiHf Pareto front is consistent—the ML developed in this work did not push the combined $\bar{T}$ and $\Delta T$ performance beyond the Pareto front that was established by the physically reasoned, unpublished NiTiHf alloy developments, indicating that the New NiTiHf Pareto Front is likely a physical bound on what can be achieved via NiTiHf metallurgy.

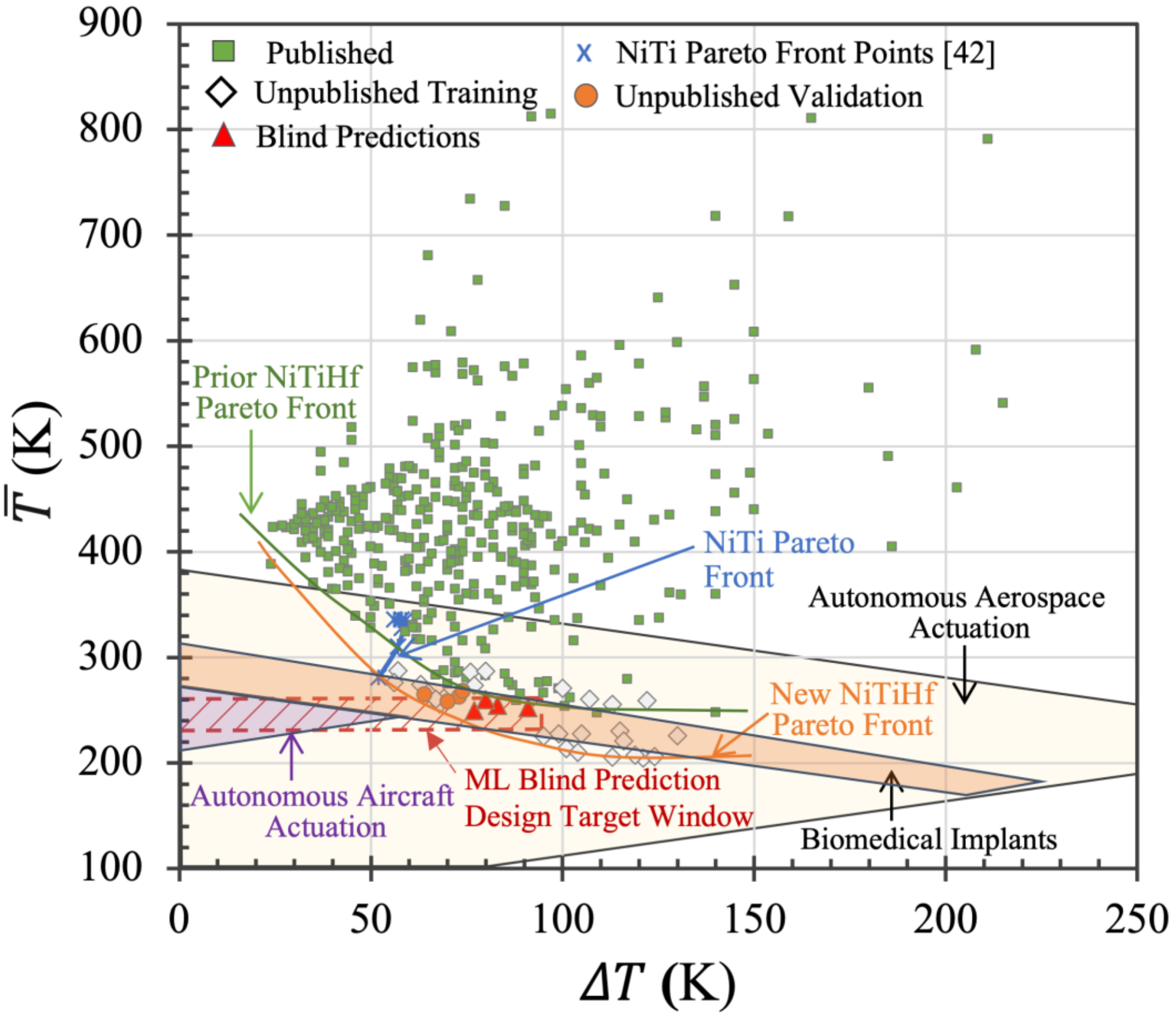

**Figure 5.** An Ashby plot shows transformation temperatures $\bar{T}$ against relative hysteresis $\triangle T$ for published Ni-Ti-Hf Pareto Front, lab unpublished work and new predictions of alloys design. The NiTi Pareto Front was obtained from [42]. Autonomous Aircraft Actuation targets actuation between takeoff and cruise. Autonomous Aerospace Actuation targets actuation between lunar day/night cycles. Biomedical Implants target a range of applications.

Still, ML-driven design showed tremendously improved accuracy. Specifically, recall that the unpublished alloy development goal was motivated by biomedical performance metrics. Only ~ 1/3 of the alloys synthesized using our best physical intuition and decades of experience met the thermal requirements—nearly 2/3 of the white diamond markers lie outside of the target region.

Contrarily, all ML predictions fall within this region. This improved accuracy represents a tremendous savings in alloy development efforts. Without ML, 2 out of every 3 synthesized alloys would have failed to meet the design goals, and with ML, all succeed (acknowledging that 5 is a limited number). Furthermore, the ML-driven design goal—alloys with minimum hysteresis at a given mean transformation temperature—was validated despite the gap that existed at the Pareto front from the unpublished, non-ML informed effort. ML identifies alloys that meet a combined $\bar{T}$ and $\Delta T$ performance target that fills that gap.

In summary, a physics-informed feature engineering approach for multi-step heat treatment schedules has been developed. Just as the physical thermodynamic and kinetic relationships (e.g., JMAK growth kinetics) have guided metallurgy for more than 80 years, a data processing methodology for ML-driven alloy design is proposed, one that is useful for any alloy system that undergoes phase transformations when heat treated. In more than 70 years of physics-based modeling developments, we still do not have a quantitative, broadly applicable, quantitative model for thermal hysteresis of precipitate strengthened SMAs. This physics-informed approach is shown to be the key to model the effects of metallurgical processing variations across multiple processing stages on mean transformation temperatures and thermal hysteresis of SMAs. At a high level, this demonstrates the ability to develop robust materials informatics approaches that work on limited data volumes, data that are very expensive and take decades to generate, rather than relying on computational or high-throughput materials databases containing many thousands of observations. The feature engineering innovation demonstrated here enables an ML CPP model able directly inform manufacturing.

**Acknowledgments:** This project was performed within the Alliance for the Development of Additive Processing Technologies (ADAPT) Center and funded by the Department of Defense, Office of Economic Adjustment, Defense Manufacturing Industry Resilience Program, Michael Gilroy program manager, grants #CTGG1 2016-2166 and #ST1605-19-03. Malcolm Davidson and Chris Borg of Citrine Informatics assisted with publishing the shape memory alloys database on Citrination. O.B. acknowledges support from the NASA Transformative Aeronautics Concepts Program, Transformational Tools & Technologies (TTT) Project.

**Competing interests:** Authors declare no competing interests.

**Data and materials availability:** The data are available on Supplementary Materials and Citrination as “NiTiHf Shape Memory Alloys,” Citrination, 2018. Available at https://citrination.com/teams/45/resources.

**List of Supplementary Materials:**
Figs. S1 to S7
Tables S1 to S2
Reference

**Supplementary Materials**

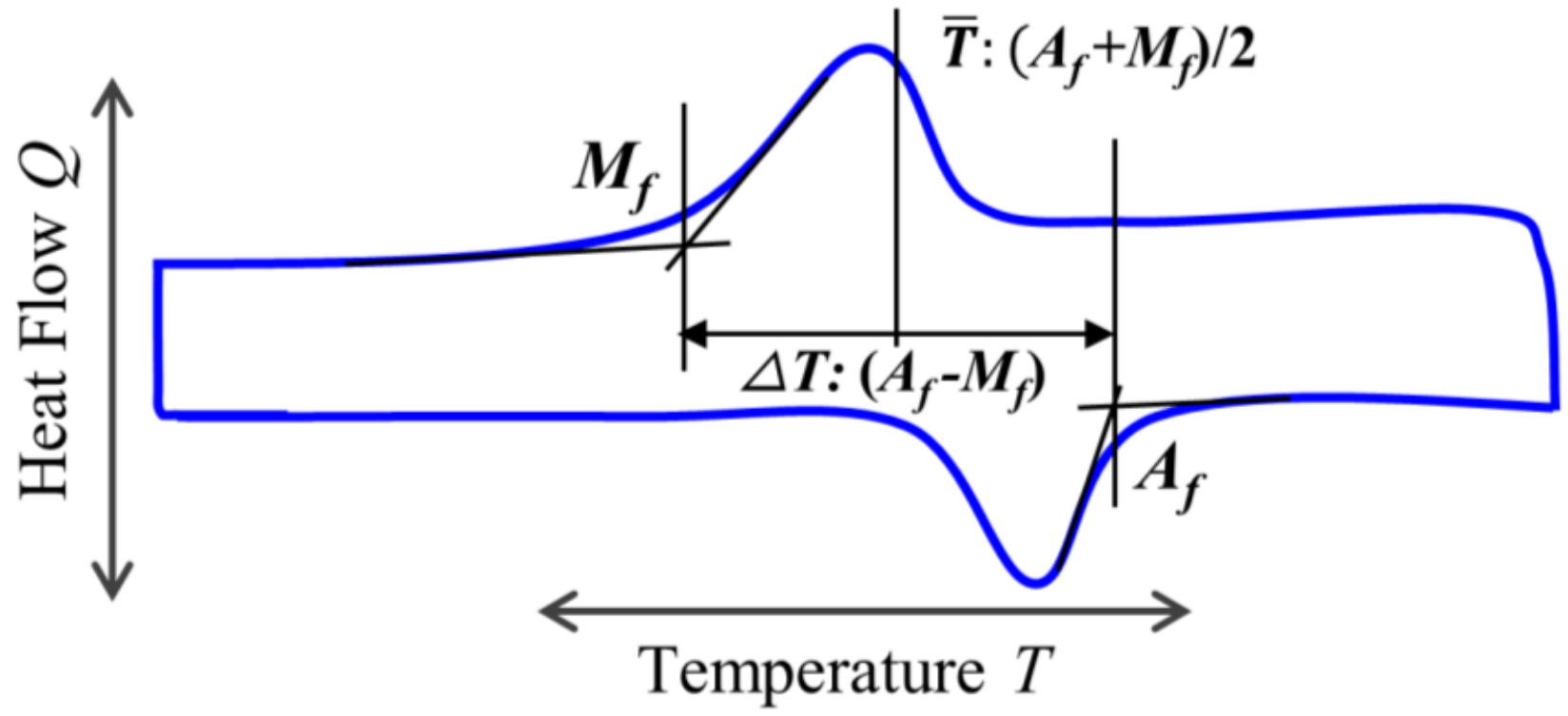

**Fig. S1. Schematic of DSC (Differential scanning calorimetry) data analysis to determine $\overline{T}$ and $\Delta T$.**

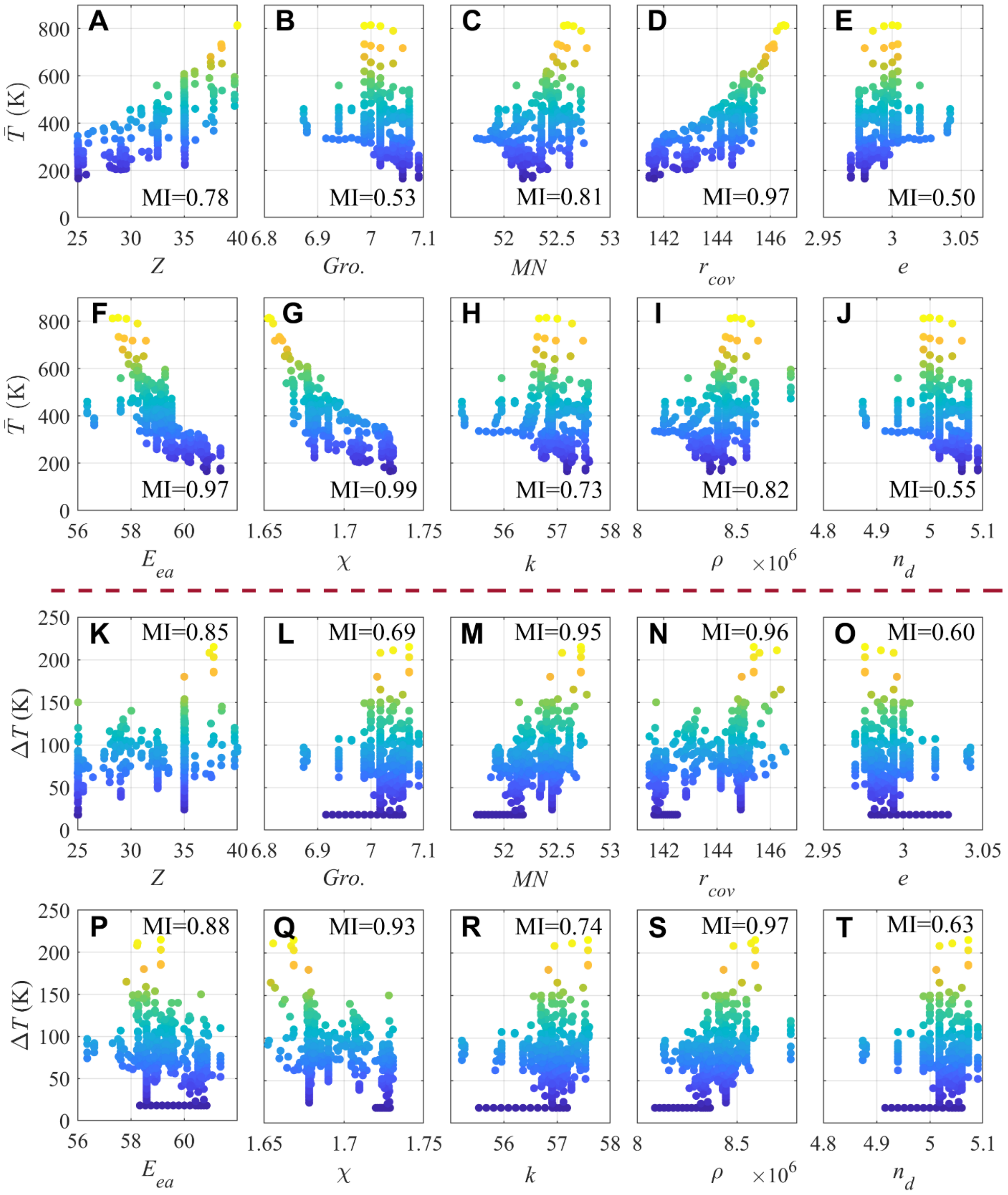

**Fig. S2. 2D scatterplots of chemical composition-based input features vs. each calculated output** (A-J) $\overline{\boldsymbol{T}}$ **and** (K-T) $\triangle \boldsymbol{T}$**.** Lighter colors indicate higher output values and vice versa. The MI scores with respect to the property are also inserted. Comparting the MI scores with the trends demonstrates that high MI score does not necessarily indicate a strong, direct correlation. For example, the correlative trend in N is not nearly as strong or obvious as the trend in A. The features

provide a general and relatively simple representation that reflect physical and chemical aspects of contributions for predicting alloys properties. For example, valence $e$ (panels E, O) equals the number of electrons gained, lost, or shared in order to form the stable compounds. Pauling electronegativity $\chi$ (panel G, Q) describes tendencies in chemical bonding. The strong chemical bonding gives rise to large resistance to shape/volume change, and high bulk and shear moduli. The elastic modulus of parent phase influences the transformation temperature [83]. Larger elastic modulus of the parent phase, cooling should continue before critical temperature point is reached; therefore, the $\bar{T}$ is depressed and vice versa. The atomic radius $r_{cal}$ or $r_{cov}$ (panels D, N) have been shown to influence the thermal hysteresis $\triangle T$ [84]. Valence electrons from $d$ orbital $n_d$ (panels J, T) accounts for most variations of total valence electrons $n$. $d$ orbital electrons count is a powerful tool for understanding the chemistry of transition metal complexes [85,86].

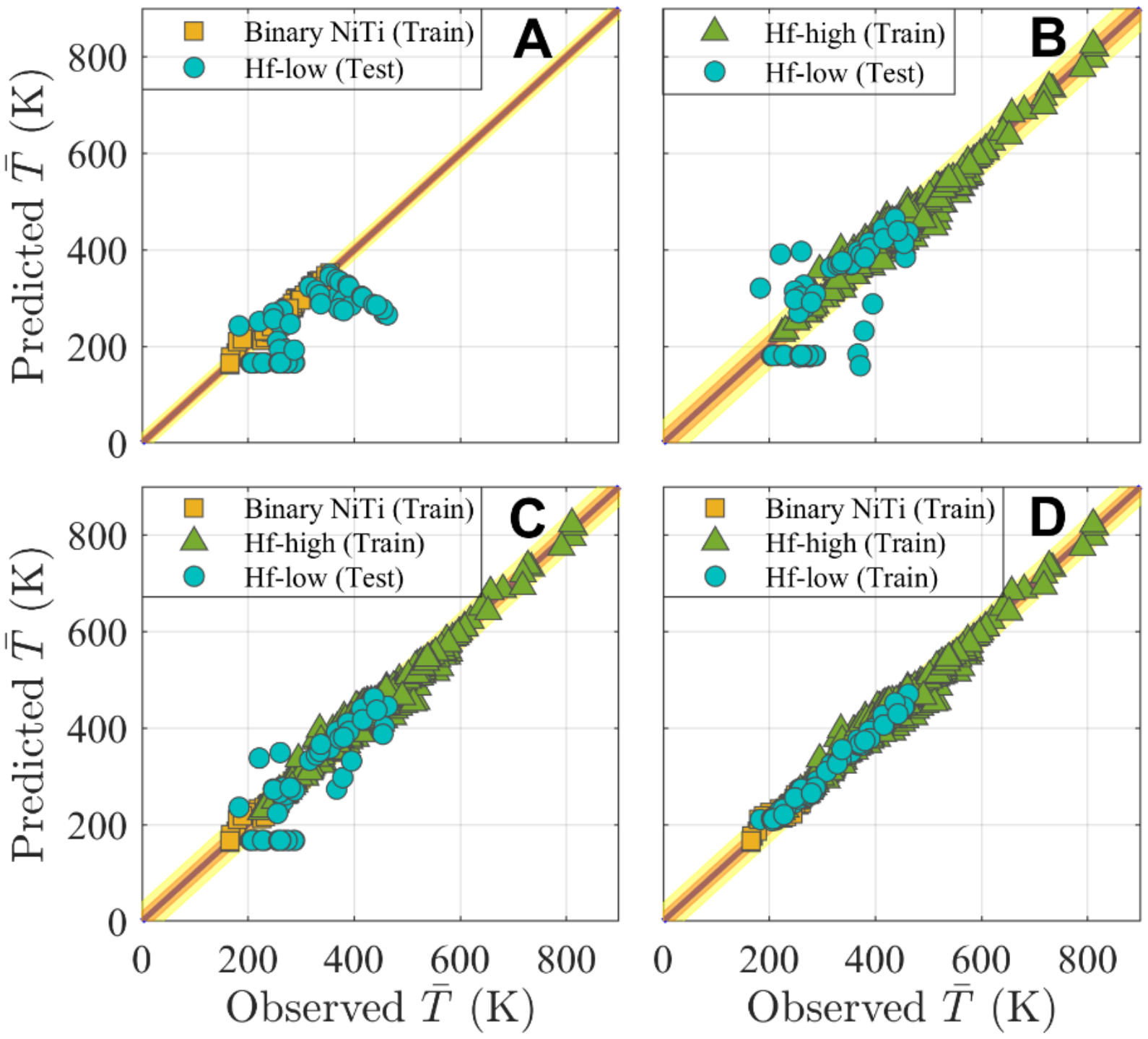

**Fig. S3. Cross-validation results for $\bar{T}$ models trained on subsets of the database, then tested on target Hf-low alloys data.** (A) binary NiTi, (B) Hf-high, and (C) binary NiTi + Hf-high data. (D) The model trained with mixed family data source. Overall, each model performed well on datasets belonging to their training subset, but not as well on the Hf-low test data (in A, B, C). The model trained on binary NiTi subset (A) performs better for lower $\bar{T}$ (< 400 K) whereas the model trained on Hf-high family (B) performs better for higher $\bar{T}$ (> 400 K). The model trained with binary NiTi and Hf-high datasets (C) shows improved performance in predicting the Hf-low test data comparatively. However, the model trained using all three data sources together (D) achieves $R^2$=0.98 and lower predicted uncertainty $\sigma$ = 20 K. These results show that each of the composition subsets of the database provide critical statistics to the overall, final model performance for $\bar{T}$.

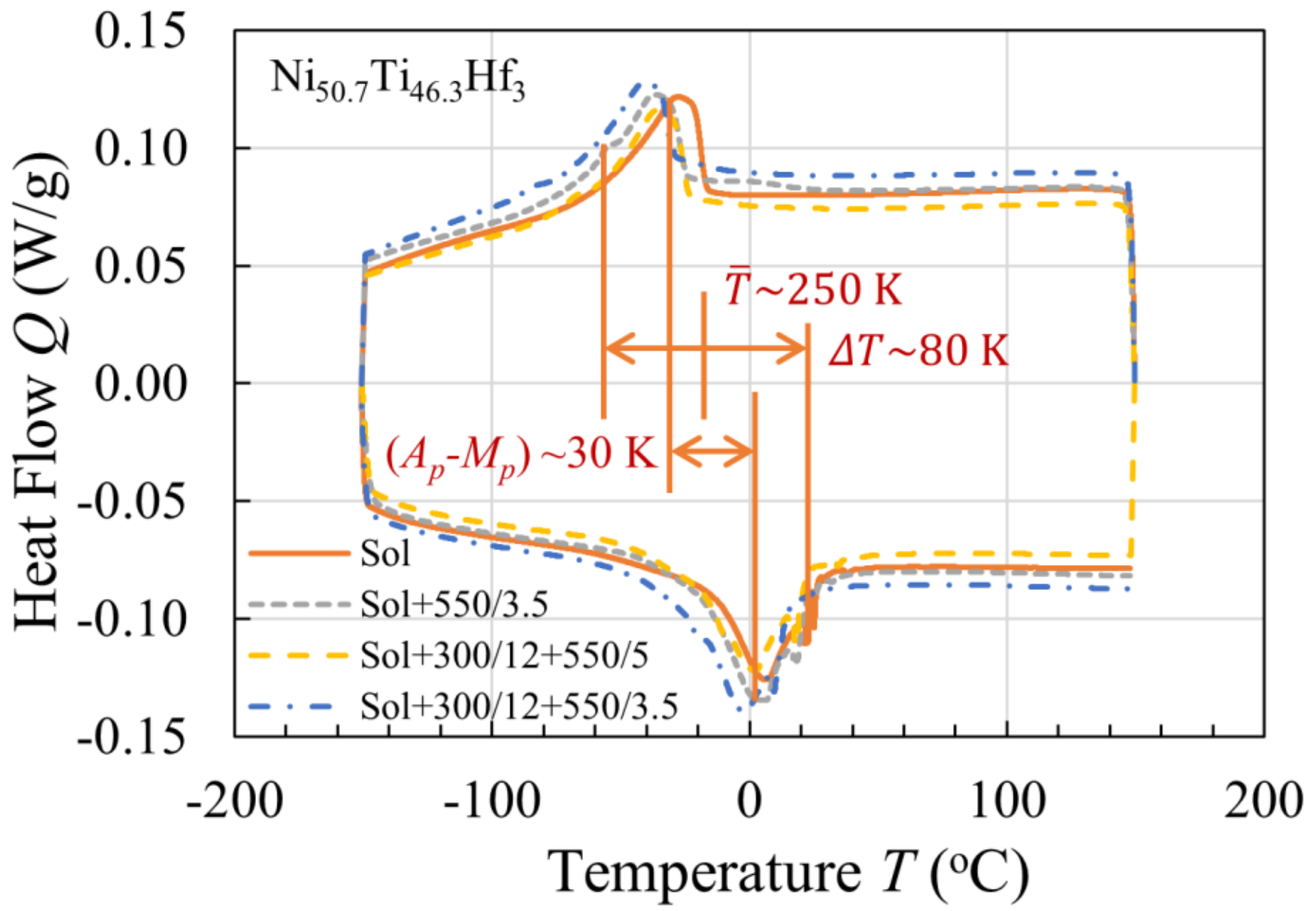

**Fig. S4. DSC measurements of the predictively designed $Ni_{50.7}Ti_{46.3}Hf_3$ alloy subjected to different heat-treatment paths**. These results show the sensitivity of one of the blind predictions to the use of heat treatment schedules other than the one used in the design. While the results are similar, they confirm that indeed the ML validation design selected the minimum hysteresis of the schedules considered by the model for this composition – the other heat treatments all show greater $\triangle T$. Generally, $\bar{T}$ was about 250 K (Fig. 4A) while $\triangle T$ varied from 80 K to just over 100 K. The peak hysteresis ($A_p$-$M_p$), as used in other works [8,42] is 30 K to 50K. These results are summarized in the last 4 rows of Table S2.

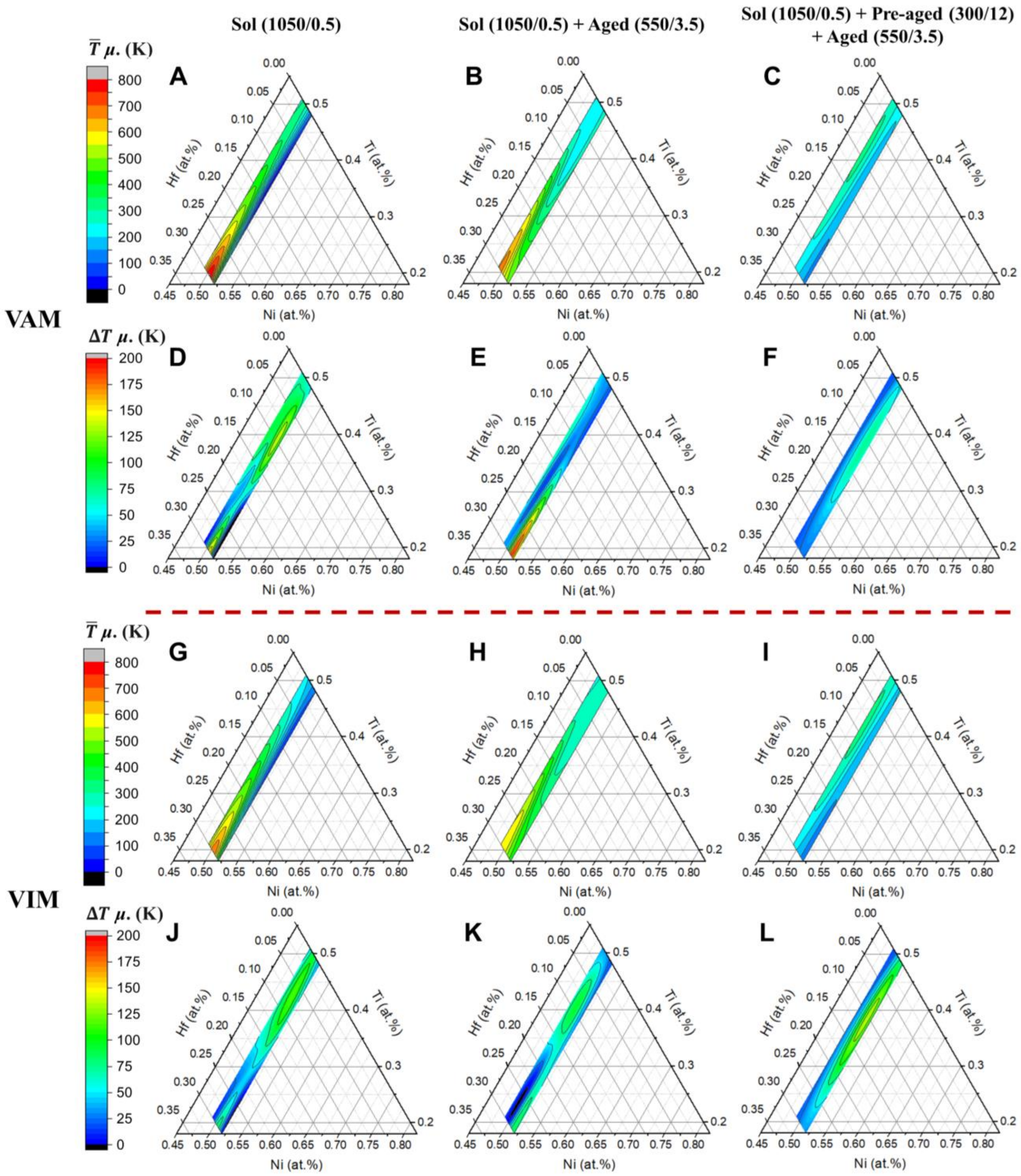

**Fig. S5. Predicted mean ($\mu$) values of $\overline{T}$ and $\triangle T$ plotted on ternary composition diagrams for different HTs and synthesis methods.** (A-F) VAM and (G-L) VIM melting methods followed by different heat-treatment schedules (A, D, G, J) Sol (1050 $^{o}$C/0.5 h, WQ), (B, E, H, K) Sol (1050 $^{o}$C/0.5 h, WQ) + Aged (550 $^{o}$C/3.5 h, AQ), and (C, F, I, L) Sol (1050 $^{o}$C/0.5 h, WQ) + Pre-aged (300 $^{o}$C/12 h, AQ) + Aged (550 $^{o}$C/3.5 h, AQ). Each prediction was constrained to $0 \leq$ Hf at.% $\leq$ 30 and $49 \leq$ Ni at.% $\leq 52$ chemistries. The predicted variances are shown Fig. S6. The alloys

demonstrate a significant composition, HTs and synthesis method dependence. Firstly, for both manufacturing methods, it presents $\bar{T}$ predicted profiles of Sol + Pre-aged + Aged HTs condition are generally lower than Sol + Aged, and Sol HTs profiles exhibit largest $\bar{T}$ value. Specifically, $\bar{T}$ profiles of Sol condition are generally 100 K higher than Sol + Aged HTs, and 300 K higher than Sol + Pre-aged + Aged HTs. Secondly, as expected, $\bar{T}$ of Hf-high content alloys are greater than those of Hf-low alloys. $\bar{T}$ keeps almost constant at Ni< 50 at.%, and then generally decreases with the Ni content increasing. Finally, the variation tendency of $\triangle T$ are more complicated and are difficult to see directly from ternary plot, a set of typical tendency curves are represented in Fig. S7. Furthermore, comparison of ternary profiles indicate that synthesis method has a strong influence on $\bar{T}$. The VIM synthesized alloys generally lower than VAM alloys by about 50-100 K. As expected, alloys property does sensitive to different synthesis ways and the ML model prediction captures the underlying phenomenon of alloys fabrication. This is because graphite crucibles are generally used for VIM whereas VAM production procedure does not need any graphite crucible. The carbon contamination TiC form during VIM solidification will increase the matrix Ni concentration, which in turns depresses $\bar{T}$ [58,59].

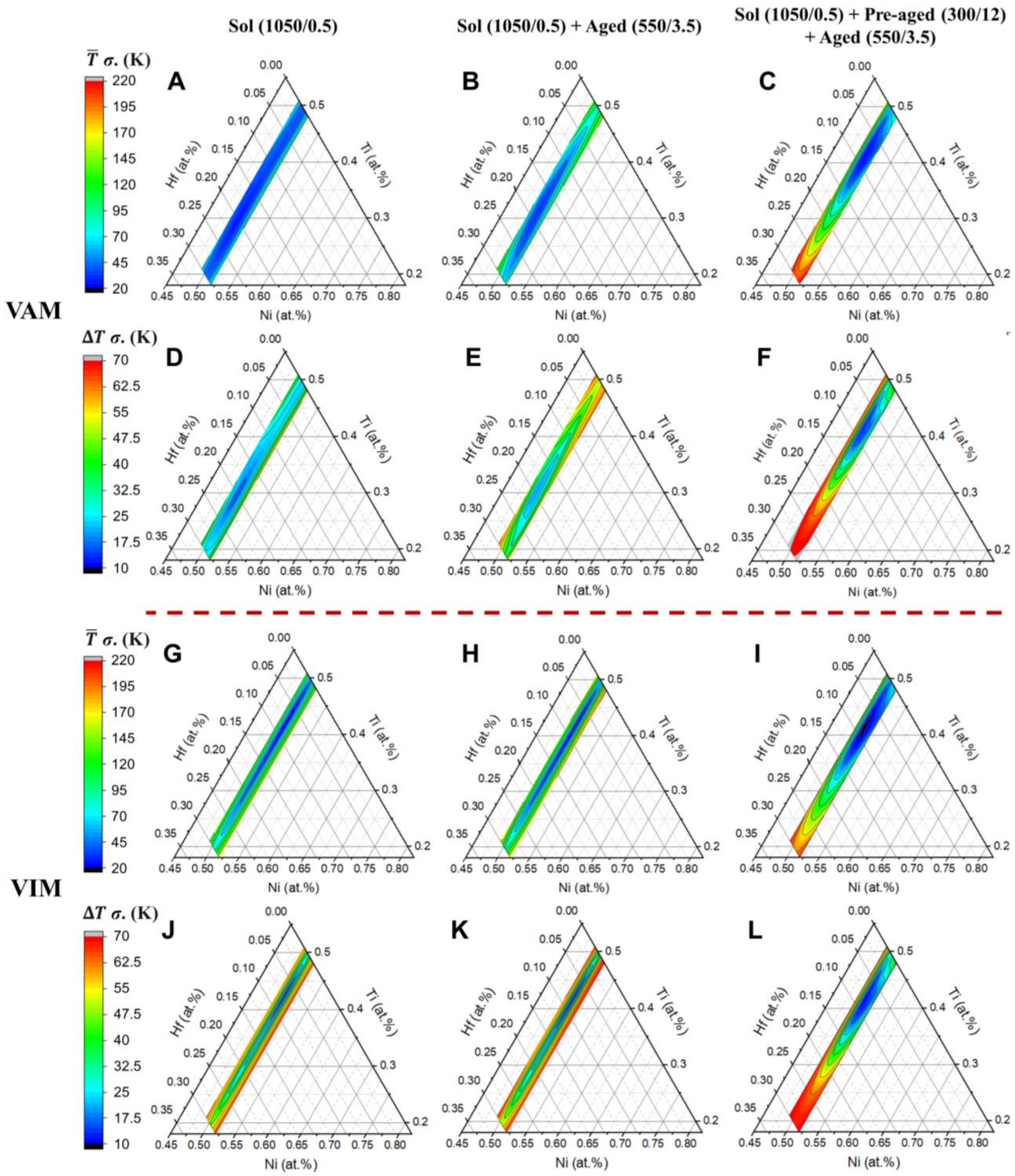

**Fig. S6. Predicted variances ($\sigma$) of $\overline{T}$ and $\triangle T$ plotted on ternary composition diagrams for different HTs and synthesis methods.** (A-F) VAM and (G-L) VIM melting methods followed by different heat-treatment schedules (A, D, G, J) Sol (1050 $^{o}$C/0.5 h, WQ), (B, E, H, K) Sol (1050 $^{o}$C/0.5 h, WQ) + Aged (550 $^{o}$C/3.5 h, AQ), and (C, F, I, L) Sol (1050 $^{o}$C/0.5 h, WQ) + Pre-aged (300 $^{o}$C/12 h, AQ) + Aged (550 $^{o}$C/3.5 h, AQ). Each prediction was constrained to $0 \leq$ Hf at.% $\leq 30$ and $49 \leq$ Ni at.% $\leq 52$ chemistries. The corresponding predicted mean values are shown Fig.

S5. It is noted that in Hf-high region of Sol + Pre-aged + Aged HTs condition, the predicted uncertainties either for $\bar{T}$ and $\triangle T$ are very large. In contrast, the uncertainties in this Hf-high region for Sol and Sol + Aged conditions are largely depressed. This is because there are no pre-aged datasets with Hf-high content within the database.

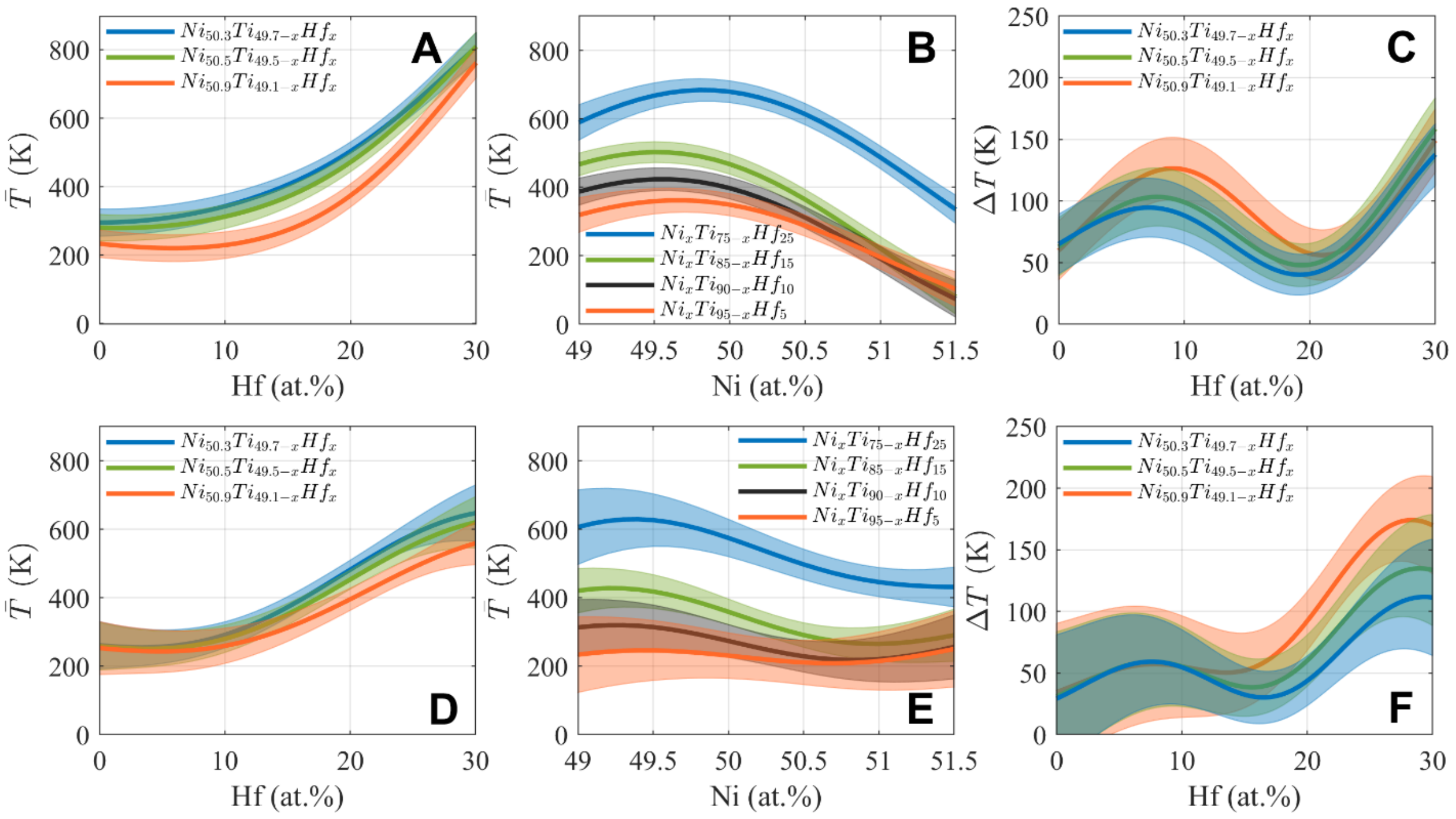

**Fig. S7. 2D model prediction plots show compositional and HTs dependencies.** Predictive tendency curves extracted from predictive ternary profiles of under (A-C) Sol and (D-F) Sol + Aging process conditions. (A, D) The variation of $\bar{T}$ with Hf content for various selected Ni contents; (B, E) variation of $\bar{T}$ with Ni content for different Hf contents; and (C, F) relative hysteresis $\triangle T$ variations against Hf content change.

**Table S1. A summary of the 26 previously unpublished process-property datasets for Hf-low alloys that were used in ML model training and testing.**

| Ni (at. %) | Ti (at. %) | Hf (at. %) | Synthesis method | Homogenization (ºC/h) | Pre-aging (ºC/h) | Final aging (ºC/h) | $M_f$ (ºC) | $M_s$ (ºC) | $A_s$ (ºC) | $A_f$ (ºC) |
|---|---|---|---|---|---|---|---|---|---|---|
| 51.5 | 42.5 | 6 | VIM | 1050/0.5 | 300/12 | 550/13.5 | -110 | -85 | -29 | -9 |
| 50.3 | 43.7 | 6 | VIM | 1050/0.5 | 23/0 | 23/0 | -110 | -71 | -39 | 6 |
| 50.3 | 43.7 | 6 | VIM | 1050/0.5 | 23/0 | 550/3.5 | -39 | -22 | 11 | 26 |
| 50.3 | 43.7 | 6 | VIM | 1050/0.5 | 300/12 | 550/7.5 | -25 | -14 | 19 | 31 |
| 50.3 | 43.7 | 6 | VIM | 1050/0.5 | 300/12 | 550/13.5 | -51 | -38 | 1 | 17 |
| 51.5 | 42 | 6.5 | VIM | 1050/0.5 | 300/12 | 550/13.5 | -115 | -89 | -31 | -11 |
| 51.5 | 41.5 | 7 | VIM | 1050/0.5 | 300/12 | 550/13.5 | -124 | -94 | -32 | -11 |
| 50.3 | 41.7 | 8 | VIM | 1050/0.5 | 300/12 | 550/0.5 | -100 | -61 | -35 | 15 |
| 50.3 | 41.7 | 8 | VIM | 1050/0.5 | 300/12 | 550/7.5 | -14 | -3 | 29 | 43 |
| 50.3 | 41.7 | 8 | VIM | 1050/0.5 | 300/12 | 550/13.5 | -30 | -16 | 14 | 33 |
| 51 | 41 | 8 | VIM | 1050/0.5 | 300/12 | 550/7.5 | -129 | -94 | -25 | -8 |
| 51 | 41 | 8 | VIM | 1050/0.5 | 300/12 | 550/13.5 | -95 | -65 | -17 | 0 |
| 50.3 | 41.7 | 8 | VIM | 1050/0.5 | 23/0 | 23/0 | -66 | -32 | -17 | 41 |
| 50.3 | 41.7 | 8 | VIM | 1050/0.5 | 23/0 | 300/12 | -74 | -36 | -21 | 39 |
| 50.3 | 41.7 | 8 | VIM | 1050/0.5 | 23/0 | 550/3.5 | -52 | -6 | -33 | 48 |
| 50.3 | 41.2 | 8.5 | VIM | 1050/0.5 | 23/0 | 550/3.5 | -75 | -8 | -19 | 47 |
| 51 | 40.5 | 8.5 | VIM | 1050/0.5 | 300/12 | 550/7.5 | -129 | -95 | -21 | -5 |
| 51 | 40.5 | 8.5 | VIM | 1050/0.5 | 300/12 | 550/13.5 | -95 | -59 | -13 | 4 |
| 50.3 | 41.2 | 8.5 | VIM | 1050/0.5 | 300/12 | 550/0.5 | -112 | -70 | -34 | 18 |
| 50.3 | 41.2 | 8.5 | VIM | 1050/0.5 | 300/12 | 550/7.5 | -25 | 2 | 18 | 51 |
| 50.3 | 41.2 | 8.5 | VIM | 1050/0.5 | 300/12 | 550/13.5 | -38 | -8 | 11 | 39 |
| 51 | 40 | 9 | VIM | 1050/0.5 | 300/12 | 550/7.5 | -125 | -91 | -21 | -6 |
| 51 | 40 | 9 | VIM | 1050/0.5 | 300/12 | 550/13.5 | -98 | -60 | -13 | 7 |
| 50.3 | 40.7 | 9 | VIM | 1050/0.5 | 300/12 | 550/7.5 | -46 | -30 | 9 | 23 |
| 50.3 | 40.7 | 9 | VIM | 1050/0.5 | 300/12 | 550/13.5 | -46 | -32 | 6 | 21 |
| 50.3 | 40.7 | 9 | VIM | 1050/0.5 | 23/0 | 550/3.5 | -26 | -4 | 24 | 54 |

**Table S2. A summary of the 17 datasets used to test the "blind prediction" capability of the ML models.**

| Ni (at. %) | Ti (at. %) | Hf (at. %) | Synthesis method | Homogenization (ºC/h) | Pre-aging (ºC/h) | Final aging (ºC/h) | $M_f$ (ºC) | $M_s$ (ºC) | $A_s$ (ºC) | $A_f$ (ºC) |
|---|---|---|---|---|---|---|---|---|---|---|
| 50.3 | 43.7 | 6 | VIM | 1050/0.5 | 300/12 | 550/3.5 | -40 | -22 | 12 | 24 |
| 50.3 | 41.7 | 8 | VIM | 1050/0.5 | 300/12 | 550/3.5 | -47 | -29 | 8 | 26 |
| 50.3 | 41.2 | 8.5 | VIM | 1050/0.5 | 300/12 | 550/3.5 | -42 | -25 | 16 | 32 |
| 50.3 | 40.7 | 9 | VIM | 1050/0.5 | 300/12 | 550/3.5 | -50 | -34 | 5 | 20 |
| 50 | 47 | 3 | VAM | 1050/0.5 | 300/12 | 550/3.5 | 23 | 57.5 | 64 | 105 |
| 50.4 | 46.6 | 3 | VAM | 1050/0.5 | 300/12 | 550/3.5 | -22 | 12 | 15 | 61 |
| 50.5 | 38.5 | 11 | VAM | 1050/0.5 | 300/12 | 550/3.5 | -21 | 25 | 44 | 83 |
| 50.4 | 37.6 | 12 | VAM | 1050/0.5 | 300/12 | 550/3.5 | 18 | 39 | 69 | 100 |
| 50.7 | 46.3 | 3 | VAM | 1050/0.5 | 300/12 | 550/3.5 | -62 | -28 | -25 | 15 |
| 50.7 | 46.3 | 3 | VAM | 1050/0.5 | 300/12 | 550/5 | -60 | -22 | -26 | 23 |
| 50.7 | 46.3 | 3 | VAM | 1050/0.5 | 23/0 | 550/3.5 | -66 | -23 | -18 | 25 |
| 50.7 | 46.3 | 3 | VAM | 1050/0.5 | 23/0 | 23/0 | -53 | -16 | -14 | 27 |